\PassOptionsToPackage{table}{xcolor}

\documentclass[11pt]{article}

\usepackage[final]{acl}

\usepackage{times}
\usepackage{latexsym}
\usepackage{xcolor}

\usepackage[T1]{fontenc}

\usepackage[utf8]{inputenc}

\usepackage{microtype}

\usepackage{inconsolata}

\usepackage{graphicx}

\usepackage{algorithm}
\usepackage{algpseudocode}
\usepackage{amsmath}
\usepackage{amssymb}
\usepackage{booktabs}
\usepackage{titletoc}
\usepackage{pifont}
\usepackage{subcaption}
\usepackage{multirow}
\usepackage{listings}
\usepackage{enumitem}
\usepackage{xcolor}
\usepackage{tabularx}
\usepackage{array}
\usepackage{ragged2e}
\usepackage{url}
\usepackage[most]{tcolorbox}

\newcommand{\promptrole}[1]{\textbf{#1:}}
\newtcblisting{promptbox}{%
  listing only, breakable, enhanced,
  colback=black!6, colframe=black!75,
  boxrule=0.8pt, arc=2mm, outer arc=2mm,
  left=6pt, right=6pt, top=6pt, bottom=6pt,
  before skip=6pt, after skip=10pt,
  listing options={%
    basicstyle=\small\rmfamily,
    columns=fullflexible, keepspaces=true,
    breaklines=true, breakatwhitespace=false, breakindent=0pt,
    showstringspaces=false,
    escapeinside={(*@}{@*)},
    literate={`}{\textasciigrave}1 {'}{\textquotesingle}1 {-}{{-}}1
             {*}{\textasteriskcentered}1,
  }%
}

\title{CR-Bench-v2: Evaluating the Real-World Utility of AI Code Review Agents}

\author{Kristen Pereira\thanks{Equal contribution.}, Neelabh Sinha\footnotemark[1], Rajat Ghosh\footnotemark[1], Debojyoti Dutta \\
  Nutanix, Inc. \\
  \texttt{\{kristen.pereira, neelabh.sinha, rajat.ghosh, debojyoti.dutta\}@nutanix.com}}

\begin{document}
\maketitle
\begin{abstract}

The rapid growth of automated code generation has created a pressing need for automated code review. Yet code review resists full automation: it depends on context beyond the \textit{diff}, on the author's intent, and on judgments about design and maintainability on which even expert reviewers disagree. Existing code review benchmarks lack standardization and rely largely on noisy human review comments scored with text-similarity metrics, which reward surface overlap and fail to measure precision. We introduce \texttt{CR-Bench}, a defect-centric evaluation framework that recasts SWE-Bench issues as pull-request review tasks with full repository context. After filtering for review-detectable defects and grounding each to a single originating pull request, \texttt{CR-Bench} contains 584 instances, with a 174-instance \texttt{CR-Bench-Verified} subset derived from SWE-Bench Verified. Each instance pairs a pull request with a known defect, its ground-truth fix, and a taxonomy label for bug category, impact, and severity. We propose two complementary metrics: \emph{Known-Defect Detection Rate} (KDDR), which measures whether an agent identifies the labeled defect, and \emph{Usefulness Rate}, which measures the fraction of review comments that are actionable. We evaluate eight code review agents over four backbones, plus commercially available Gemini Code Assist, using three independent LLM judges. Even the strongest systems detect only about one-third of known defects (KDDR $\approx$ 33\%), and Gemini Code Assist detects 8.1\%. 

\end{abstract}

\section{Introduction}
\label{sec:intro}

% \begin{figure}[htbp]
%     \centering
    
%     \begin{subfigure}[t]{0.32\linewidth}
%         \centering
%         \includegraphics[width=\linewidth]{assets/PR.pdf}
%         \caption{Pull Request ($PR$)}
%         \label{fig:pr}
%     \end{subfigure}
%     \hfill
%     \begin{subfigure}[t]{0.32\linewidth}
%         \centering
%         \includegraphics[width=\linewidth]{assets/patch.pdf}
%         \caption{Patch ($r_\delta$)}
%         \label{fig:patch}
%     \end{subfigure}
%     \hfill
%     \begin{subfigure}[t]{0.32\linewidth}
%         \centering
%         \includegraphics[width=\linewidth]{assets/comment.pdf}
%         \caption{Review Comment ($r_c$)}
%         \label{fig:comment}
%     \end{subfigure}

%     \caption{Example of an instance from \texttt{CR-Bench} dataset: In a pull request, a comment should be made to address the specified bug. The fix is removing the lines of code marked in \textcolor{red}{RED} and adding the ones marked in \textcolor{green}{GREEN}.}
%     \label{fig:example}
% \end{figure}

 Despite significant commercial interest~\citep{coderabbit2024,googlecodeassist2024,claudecode2024}, code review automation remains difficult~\citep{modern-code-review-survey,modern-code-review-challenges} because it is time-consuming for large multi-file repositories, costly, and inherently subjective~\citep{github2026codereview, swr-bench,harnessing-llm-codereview}. Existing code review benchmarks (Table~\ref{tab:novelty-matrix}, Appendix) fail in pull requests with at least one of these scenarios: multiple diff hunks, multi-file PR context, systematic categorization, and verifiable defect handling~\citep{swr-bench,harnessing-llm-codereview,trans-review-data,t5-codereview,autotransform}.

Feedback from engineers using review agents on live repositories shows that noisy comments hamper productivity and slow adoption. Before deploying an agent, a team needs to know how often it catches real defects and how much of its output is worth reading.

In this paper, we introduce \texttt{CR-Bench/CR-Bench-Verified}, a benchmark for evaluating how well code review agents detect defects and generate useful review comments. Each review task consists of the source repository, the identification number of the pull request (PR) within that repository, a description of the defect introduced by the PR, the ground-truth patch that resolves the defect, and three taxonomy labels: category, impact, and severity (see Figure~\ref{fig:taxonomy} in the Appendix).  It prioritizes defect-identifying reviews that target functional, performance, reliability, or security regressions: defects cause system faults and can be verified, whereas stylistic preferences are subjective and better handled by linters or human judgment. 

Our key contributions are as follows:
\begin{itemize}[leftmargin=*, itemsep=0.1em, topsep=0.1em]
\item \textbf{\texttt{CR-Bench} and \texttt{CR-Bench-Verified}.} A benchmark of real-world, detectable defects for evaluating automated code review agents, with each defect labeled under a fine-grained taxonomy of category, impact, and severity.
\item \textbf{Two-axis evaluation metrics.} A measurement framework that scores code review agents on both logical defect detection and the usefulness of their generated comments.
\item \textbf{Comprehensive benchmarking.} An evaluation of eight code review agents and one commercial code review agent, showing that even the strongest agent detects only about one-third of known defects, with practical guidance for teams deploying review agents (\S~\ref{sec:conclusion}).
\end{itemize}

\section{Related Work}
\label{sec:related_work}

Table~\ref{tab:novelty-matrix} (Appendix) compares \texttt{CR-Bench} with prior code review datasets on four properties: multiple diff hunks per task, full Pull Request (PR) context, categorization of the reviewed issue, and a focus on a verifiable defect. It also lists how each dataset scores reviews.

\textbf{Learning-based code review.} Early work frames review automation as a sequence-to-sequence task over a single code change. \citet{trans-review-data} and \citet{t5-codereview} train models to generate reviewer comments and revised code for method-level changes, and AutoTransform~\citep{autotransform} learns the code transformations that reviewers request. CodeReviewer~\citep{codereviewer} pre-trains on diff hunks for review comment generation and code refinement. Each task pairs one change with a reference comment or revision, so these datasets lack multi-hunk and cross-file PR context and carry no categorization of the underlying issue. They are scored with Exact Match, BLEU, or CodeBLEU, which reward similarity to the reference rather than detection of a defect. LLM agents such as CodeAgent~\citep{codeagent} and production tools such as CodeRabbit~\citep{coderabbit2024}, Gemini Code Assist~\citep{googlecodeassist2024}, and Qodo~\citep{qodopragent2024} now review entire PRs, which calls for repository-scale evaluation.

\textbf{Code review benchmarks.} SWR-Bench~\citep{swr-bench} provides full PR context at repository scale, includes multiple diff hunks, categorizes its PRs, and scores comments with an LLM judge~\citep{g-eval} using precision, recall, and F1. However, it scores hits against human-verified change points, so its ground truth inherits the subjectivity of those changes and is not tied to a verifiable defect. \texttt{CR-Bench/CR-Bench-Verified} keep the multi-hunk, full-PR setting, ground each instance in a SWE-Bench defect~\citep{swe-bench} labeled by category, impact, and severity, and measure both defect detection (KDDR) and developer utility (Usefulness Rate). To the best of our knowledge, they are the first code review benchmarks with all four properties in Table~\ref{tab:novelty-matrix}.

% \textbf{Consensus-based Evaluation and Annotator Agreement.} Classical metrics like Cohen's $\kappa$~\citep{cohen_kappa} assess annotation reliability using majority voting. However, recent frameworks~\citep{dumitrache2021empirical} argue that annotator disagreement reflects inherent ambiguity rather than simple noise, advocating for explicitly modeling this instead of collapsing it into a single consensus label. Concurrently, LLM-as-a-judge~\citep{g-eval} paradigm has created a need for rigorous criteria to determine when LLMs can replace human annotators. Addressing this, the Alternative Annotator Test (Alt-Test)~\citep{alt_test} compares an LLM's alignment with human consensus against individual human annotators. 

% \input{content/problem_definition}
\section{CR-Bench: Benchmarking Framework for Code Review}
\label{sec:dataset_overview}

 \textit{Code Review} starts with a \textit{Pull Request} ($PR$) on a base code state $(\mathcal{S_B})$ --- a proposal to integrate a sequence of code modifications into a target codebase. Formally, $PR$ is a set of $\{\mathcal{C}, \Delta, \mathcal{M}\}$, where $\mathcal{C} = \{c_1, c_2, \dots, c_n\}$ is a set of discrete \textit{commits}, $\Delta$ is the cumulative code diff, and $\mathcal{M}$ denotes the associated metadata (e.g., title, description, and author). The output for a code review process is a set of $m$ reviews: $\mathcal{R} = \{ (r_{c}, r_{\delta})_i \mid 1 \le i \le m \}$. $r_{c}$ is a natural language comment, which identifies specific defects or suggests improvements, and $r_{\delta}$ provides a concrete code-level resolution to the identified defects. The code review task is formalized in Equation \ref{eq:def}.

\begin{equation}
    (PR, \mathcal{S_B})  \xrightarrow{\text{CR Agent or Human}} \mathcal{R}^{m}
    \label{eq:def}
\end{equation}

% \textbf{Bias-Variance Trade-off.} Code review tasks often involve substantial subjectivity~\citep{swr-bench}. A given change can also have multiple valid resolutions, which increases with the size and complexity of $(PR, \mathcal{S_B})$. This can introduce review noise, making reliable automation challenging. This reflects a pronounced bias--variance trade-off that represents a central design challenge for automated code review systems. 

% Full-width float: the diagram is 3:1, so at \columnwidth it rendered at 24% of native
% size and its internal labels were unreadable. A band gives it 50%, the shape it was drawn for.
\begin{figure*}[t]
    \centering
    \includegraphics[width=\linewidth]{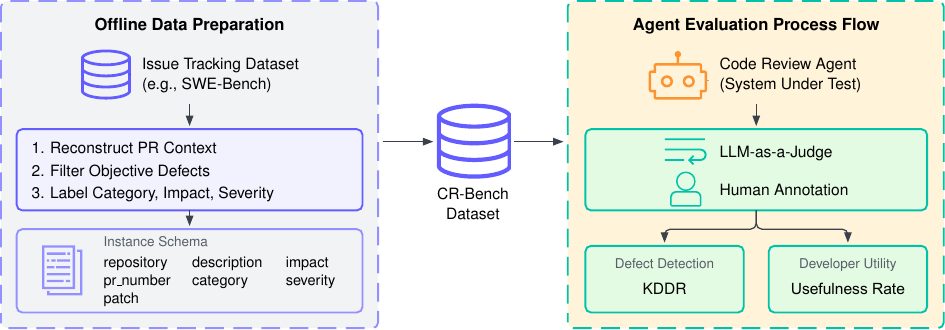}
    \caption{\textbf{\texttt{CR-Bench} framework.} Offline preparation turns SWE-Bench defects into PR-context instances with taxonomy labels. An LLM judge classifies agent comments, and human annotations validate the classifications. KDDR measures labeled-defect detection. Usefulness Rate measures developer utility.}
    \label{fig:intro}
\end{figure*}

We construct the \texttt{CR-Bench/CR-Bench-Verified} benchmarks by transforming the SWE-Bench \citep{swe-bench}/SWE-Bench Verified \citep{swe-bench-verified} datasets using Algorithm~\ref{alg:dataset_creation}.
The algorithm converts SWE-Bench's $(issue, patch)$ instances into $(PR, Review)$ instances: the PR is the earlier pull request that introduced the defect, and the reference review pairs a comment derived from the issue with the fix patch.

\subsection{Dataset Creation Algorithm}

% Algorithm 1 and Table 2 share one full-width float on purpose. As separate single-column
% floats they competed for column tops, and the algorithm alone exceeded \bottomfraction,
% so LaTeX surrendered a whole column to floats. Half of \textwidth is within 9pt of
% \columnwidth, so the pseudocode keeps its natural density here rather than stretching.
% Captions use \captionof so each keeps its own counter (Algorithm 1, Table 2).
\begin{figure*}[t]
\centering
% \vspace{0pt} makes [t] align the true top of each box rather than the first baseline,
% which otherwise dropped the algorithm ~50pt below the table beside it.
\begin{minipage}[t]{0.62\textwidth}
    \vspace{0pt}
    \hrule height 0.8pt
    \vspace{3pt}
    \captionof{algorithm}{CR-Bench Construction via SWE-Bench Transformation}
    \label{alg:dataset_creation}
    \vspace{2pt}
    \hrule height 0.4pt
    \vspace{3pt}
    \small
    \begin{algorithmic}[1]
    \Require SWE-Bench $\mathcal{S}$, LLM $\mathcal{L}$, prompts $\mathcal{P}_{D}$ (detectability), $\mathcal{P}_{C}$ (comment), $\mathcal{P}_{T}$ (tagging)
    \Ensure CR-Bench dataset $\mathcal{D}$
    \State $\mathcal{D} \gets \emptyset$
    \For{\textbf{each} instance $s \in \mathcal{S}$}
        \vspace{1pt}
        \State \textit{Detectability filtering:} $is\_detectable \gets \mathcal{L}(\mathcal{P}_{D}, s.\text{patch})$ \label{alg:dataset_creation-step_llm_classify}
        \State \textbf{continue if} $\neg is\_detectable$ \label{alg:dataset_creation-step_filter_prev}
        \vspace{1pt}
        \State \textit{PR grounding:} $c_{base} \gets s.\text{base\_commit}$
        \State $\mathcal{C}_{blame} \gets \textproc{GitBlame}(c_{base})$ \label{alg:dataset_creation-step_blame}
        \State $PR \gets \textproc{QueryGitHubAPI}(\mathcal{C}_{blame})$ \label{alg:dataset_creation-step_api}
        \State \textit{Single-PR filter:} \textbf{continue if} $|PR| \neq 1$ \label{alg:dataset_creation-step_filter_pr}
        \State $pr \gets \text{unique element of } PR$
        \vspace{1pt}
        \State \textit{Attribute generation \& aggregation:} $r_{c} \gets \mathcal{L}(\mathcal{P}_{C}, s.\text{desc})$ \label{alg:dataset_creation-step_paraphrase}
        \State $r_{\delta} \gets  s.\text{patch}$
        \State $t_{gen} \gets \mathcal{L}(\mathcal{P}_{T}, r_{\delta}, r_{c})$ \label{alg:dataset_creation-step_tagging}
        \State $\mathcal{D} \gets \mathcal{D} \cup \{ (pr, r_{\delta}, r_{c}, t_{gen}) \}$ \label{alg:dataset_creation-step_add}
    \EndFor
    \State \Return $\mathcal{D}$
    \end{algorithmic}
    \vspace{2pt}
    \hrule height 0.8pt
\end{minipage}\hfill
% Widths are 0.62/0.34 rather than an even split. The algorithm uses compact one-line guards;
% measured 1pt phase breaks preserve scanability while matching the taxonomy's height.
\begin{minipage}[t]{0.34\textwidth}
    \vspace{0pt}
    \small
    \setlength{\tabcolsep}{4pt}
    \begin{tabularx}{\linewidth}{@{}lX@{}}
    \toprule
    \multicolumn{1}{c}{\bf Taxonomy} & \multicolumn{1}{c}{\bf Elements} \\
    \midrule
    Category & Structural, IIS, RFF, Data, Concurrency, Memory, Security, Coding \\
    \midrule
    Impact & Functional Suitability, Performance Efficiency, Compatibility, Usability, Reliability, Security, Maintainability, Portability \\
    \midrule
    Severity & Low, Medium, High \\
    \bottomrule
    \end{tabularx}
    \captionof{table}{The \texttt{CR-Bench} taxonomy elements for defect labeling. Category follows~\citet{category-taxonomy-source, bug-category-DL} and impact follows ISO 25010~\citep{ISO25010}; severity is ours.}
    \label{tab:taxonomy}
\end{minipage}
\end{figure*}

Given the SWE-Bench or SWE-Bench Verified dataset $\mathcal{S}$, each instance $s \in \mathcal{S}$ is processed through three sequential phases. Every LLM call in the pipeline uses Claude Opus 4.5~\citep{claude-opus-4.5} as $\mathcal{L}$. First, we apply an LLM-based \textit{detectability filter} to classify whether the underlying issue in the patch is inferable from the code change itself. Specifically, we query an LLM $\mathcal{L}$ with a detectability prompt $\mathcal{P}_D$ (Appendix~\ref{prompt:detectable_bug}) over the patch $s.\text{patch}$, and we discard instances deemed non-detectable. This step removes cases that require external context or latent knowledge.
Second, we apply \textit{PR grounding} to link each instance to the real-world pull requests that introduced the modified code. Starting from the base commit $c_{\text{base}}$, we retrieve commit IDs of modified lines using \texttt{git blame}. Using these, we retrieve associated PRs of these commit IDs via the GitHub API. Third, we apply a \textit{single-PR filter}. We retain only the instances that map to a single PR, ensuring an unambiguous correspondence between the patch and its origin PR. Instances with zero or multiple associated PRs are discarded. Finally, we perform attribute generation and dataset aggregation. For each retained instance, we generate a natural language review comment $r_c$ using $\mathcal{L}$ with prompt $\mathcal{P}_C$ (Appendix~\ref{prompt:issue_description_paraphrase}), conditioned on the original problem description. The patch itself is retained as $r_{\delta}$. We further assign tags of category, impact, severity $t_{\text{gen}}$ using a tagging prompt $\mathcal{P}_T$ (Appendix~\ref{prompt:cateogrize_bugs}) over both the patch and generated comment. The resulting tuple $(pr, r_{\delta}, r_c, t_{\text{gen}})$ is added to the dataset $\mathcal{D}$. Together, these steps keep only actionable, context-resolvable review scenarios, and the generated attributes support fine-grained evaluation. They reduce SWE-Bench from 2,294 to 584 instances (\texttt{CR-Bench}) and SWE-Bench Verified from 500 to 174 (\texttt{CR-Bench-Verified}). The 174 \texttt{CR-Bench-Verified} tasks span 11 source repositories, led by Django with 63 (Figure~\ref{fig:histogram}, Appendix). Other repositories include mathematical libraries like sympy/sympy, and specialized tools such as astropy/astropy and scikit-learn/scikit-learn. The statistics in Table~\ref{tab:crbench_stats}, Appendix show that \texttt{CR-Bench} covers large scale PRs and an code review agents must be robust to perform well on it. An example of data instance is given in Figure~\ref{fig:example} in Appendix.

% \subsection{Taxonomy}
% \label{sec:taxonomy}

% % Taxonomy table now shares the full-width float with Algorithm 1 at the top of Sec. 4.1.

For ease of evaluation analytics, each instance of \texttt{the CR-Bench/CR-Bench-Verified} datasets is labeled into three classes: category, impact, and severity, as defined in Table \ref{tab:taxonomy} and Figure~\ref{fig:taxonomy} in the Appendix. 
In order to generate class labels, we make an LLM call (Step~\ref{alg:dataset_creation-step_tagging} in Algorithm~\ref{alg:dataset_creation}) using the problem definition ($desc$) and fix patch with a tag generation prompt $\mathcal{P^T}$ (Appendix~\ref{prompt:cateogrize_bugs}). The prompt also elaborates on the definition of each of the items in all three buckets. 

As per Figure~\ref{fig:dist_verified}, \texttt{CR-Bench-Verified} is heavily weighted toward Structural Bugs, accounting for $82.8\%$ of the corpus. These represent complex errors in logical flow. The primary impacts are Functional Suitability and Reliability. The rigor is further evidenced by $92.5\%$ of the defects being Medium or High severity. High-severity defects are predominantly linked to Reliability and Functional Suitability, representing critical failures that would likely result in system regressions if merged. The $174$ instances are what survives applying detectability filtering, PR grounding, and the single-PR filter to the $500$ human-validated instances of SWE-Bench Verified~\citep{swe-bench-verified}, a scale comparable to agentic benchmarks that verify every instance by hand, such as Terminal-Bench 2.0 with $89$ tasks~\citep{terminal-bench}.

The expansion into the full \texttt{CR-Bench} further reinforces these findings while increasing the size of the benchmark. As shown in Figure~\ref{fig:dist_full}, Appendix~\ref{app:cr_bench_stats}, this larger dataset adds instances of impacts like Maintainability, Security, and some readjustments to proportions of different categories, impacts, and bugs. This ensures that the benchmark provides a meaningful assessment of an agent's ability.

\section{Experiments}
\label{sec:experiments}

We assess the construct validity of \texttt{CR-Bench} by testing whether it reproduces well-established patterns from prior work.  We test two: iterative self-reflection on a model's own output improves over single-shot generation~\citep{reflexion}, and stronger models resolve more repository-level tasks, as reflected in the SWE-Bench leaderboard~\citep{swe-bench}. Reflexion should thus beat single-shot on each backbone, and GPT-5 backbones should beat the smaller open-weight ones.

We evaluate four backbone LLMs: two closed-API models, GPT-5.2~\citep{gpt-5,gpt-5.2} and GPT-5-mini~\citep{gpt-5}, and two open-weight models, Gemma 4 31B~\citep{gemma-4} and GPT-OSS 20B~\citep{gpt-oss}. All nine resulting systems are evaluated on \texttt{CR-Bench-Verified}. 

We evaluate each model under two harnesses of increasing complexity: a single-shot harness that issues one model call over the PR diff (Appendix~\ref{prompt:single_shot_agent}, inspired by~\citet{qodopragent2024}), and a self-reflection harness that iteratively critiques and revises its review~\citep{reflexion} (Appendix~\ref{prompt:reflexion_agent}). We also evaluate \textit{Gemini Code Assist}, a fully managed commercial agent whose underlying LLM and harness are not disclosed, by cloning PRs and repositories into a private GitHub instance and invoking it through its GitHub App\footnote{\url{https://github.com/apps/gemini-code-assist}}.

\texttt{CR-Bench} itself was built with Claude Opus 4.5 (\S\ref{sec:dataset_overview}), which we use neither as a review agent nor as a judge. We score each fixed set of reviews independently with GPT-5.4~\citep{gpt-5.4}, Claude Sonnet 4.5~\citep{claude-4.5-sonnet}, and DeepSeek V4 Flash~\citep{deepseek-v4-flash} using the same balanced evaluator prompt (Appendix~\ref{prompt:cr_verifier}).  We additionally test strict (Appendix~\ref{prompt:cr_verifier_strict}) and lenient (Appendix~\ref{prompt:cr_verifier_lenient}) prompt variants for the GPT-5.2 single-shot and Reflexion outputs.

\subsection{Evaluation Methodology with LLM-as-a-Judge}
\label{sec: cr_verifier}

\texttt{CR-Bench} uses LLM-as-a-Judge evaluators (Appendix~\ref{prompt:cr_verifier}) to classify each review comment into one of two labels: first, \textit{ Bug Hit (H)} corresponds to reviews that accurately identify or directly relate to the specific bug described in the ground truth review, and second, \textit{Valid Suggestion (S)} corresponds to reviews with constructive feedback such as stylistic improvements, performance optimizations, or edge-case handling. These reviews are significant, but not related to the primary defects in ground truth. We measure identification of the benchmark's labeled defect using the \textit{Known-Defect Detection Rate (KDDR)}. Let each of the $B$ benchmark instances contain one SWE-Bench-derived known defect. A PR may contain other unlabeled defects. Let $d_i=1$ when at least one review comment identifies instance $i$'s known defect, and $d_i=0$ otherwise:

\begin{equation}
\label{eq:kddr}
\text{KDDR} = \frac{1}{B}\sum_{i=1}^{B} d_i
\end{equation}

% KDDR names the benchmark target explicitly to avoid confusion with patch-solving accuracy.
KDDR scores review output against the defect \texttt{CR-Bench} labels for each instance, so it measures detection rather than repair and is not comparable to SWE-Bench resolve rates.

Because agents can produce useful comments beyond the single labeled defect, we also report \textit{Usefulness Rate (U)}. Here, $R$ is the total number of review comments:

\begin{equation}
\label{eq:usefulness}
U =
\frac{
H + S
}{
R
}
\end{equation}

% Results floats are declared here so that, together with the full-width taxonomy
% figure declared at the start of Results, they all fit inside the main-text page limit.
\begin{figure}[t]
    \centering
    \includegraphics[width=\columnwidth]{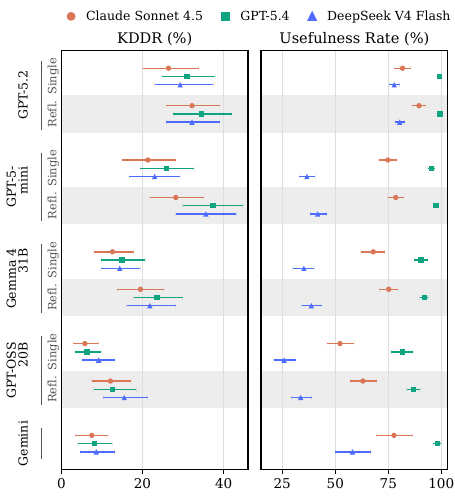}
    \caption{Judge-specific KDDR (left) and Usefulness Rate (right) for the same fixed review comments on all 174 verified instances under the balanced prompt. Single = single-shot, Refl. = Reflexion, and Gemini = Gemini Code Assist. Marker shape identifies the judge. Points are full-sample estimates. Bars are 95\% percentile confidence intervals from 1,000 instance-level bootstrap resamples.}
    \label{fig:judge_kddr_usefulness}
\end{figure}

\begin{figure}[t]
    \centering
    \includegraphics[width=\columnwidth]{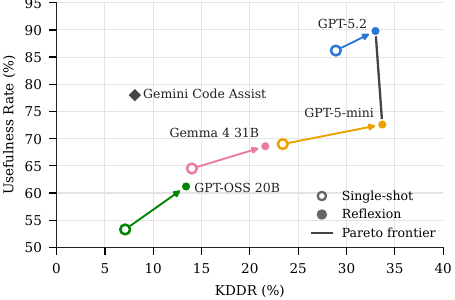}
    \caption{Three-judge mean KDDR and Usefulness Rate per system (Table~\ref{tab:results}). Arrows go from single-shot (hollow) to Reflexion (filled). The solid line is the Pareto frontier.}
    \label{fig:pareto}
\end{figure}

\begin{table}[t]
\centering
\small
\begin{tabular}{lccc}
\toprule
\textbf{Backbone} & \textbf{Claude} & \textbf{GPT-5.4} & \textbf{DeepSeek} \\
\midrule
GPT-5.2 & +5.7 & +3.4 & +2.9 \\
GPT-5-mini & +6.9$^{*}$ & +11.5$^{*}$ & +12.6$^{*}$ \\
Gemma 4 31B & +6.9$^{*}$ & +8.6$^{*}$ & +7.5$^{*}$ \\
GPT-OSS 20B & +6.3$^{*}$ & +6.3$^{*}$ & +6.3 \\
\bottomrule
\end{tabular}
\caption{Reflexion minus single-shot KDDR (percentage points) on \texttt{CR-Bench-Verified}, per backbone and judge. $^{*}$\,$p<0.05$, two-sided paired permutation test over instances. Confidence intervals and the other metrics are in Table~\ref{tab:paired_all_metrics}, Appendix~\ref{app:robustness}.}
\label{tab:paired_kddr}
\end{table}

\paragraph{Uncertainty and robustness.}
Point estimates use all 174 instances. 95\% confidence intervals come from 1,000 instance-level bootstrap resamples (seed 42) that keep each instance's comments together. Strategy comparisons are paired by \texttt{instance\_id}, with paired bootstrap intervals and two-sided permutation tests (1,000 permutations). The balanced evaluator prompt is primary; strict and lenient variants measure prompt sensitivity (Appendix~\ref{app:robustness}).

\section{Results}
\label{sec:results}

% Declared early: a full-width float lands on the page after its declaration, and
% it must stay inside the main-text page limit.
\begin{figure*}[t]
    \centering
    \includegraphics[width=0.85\linewidth]{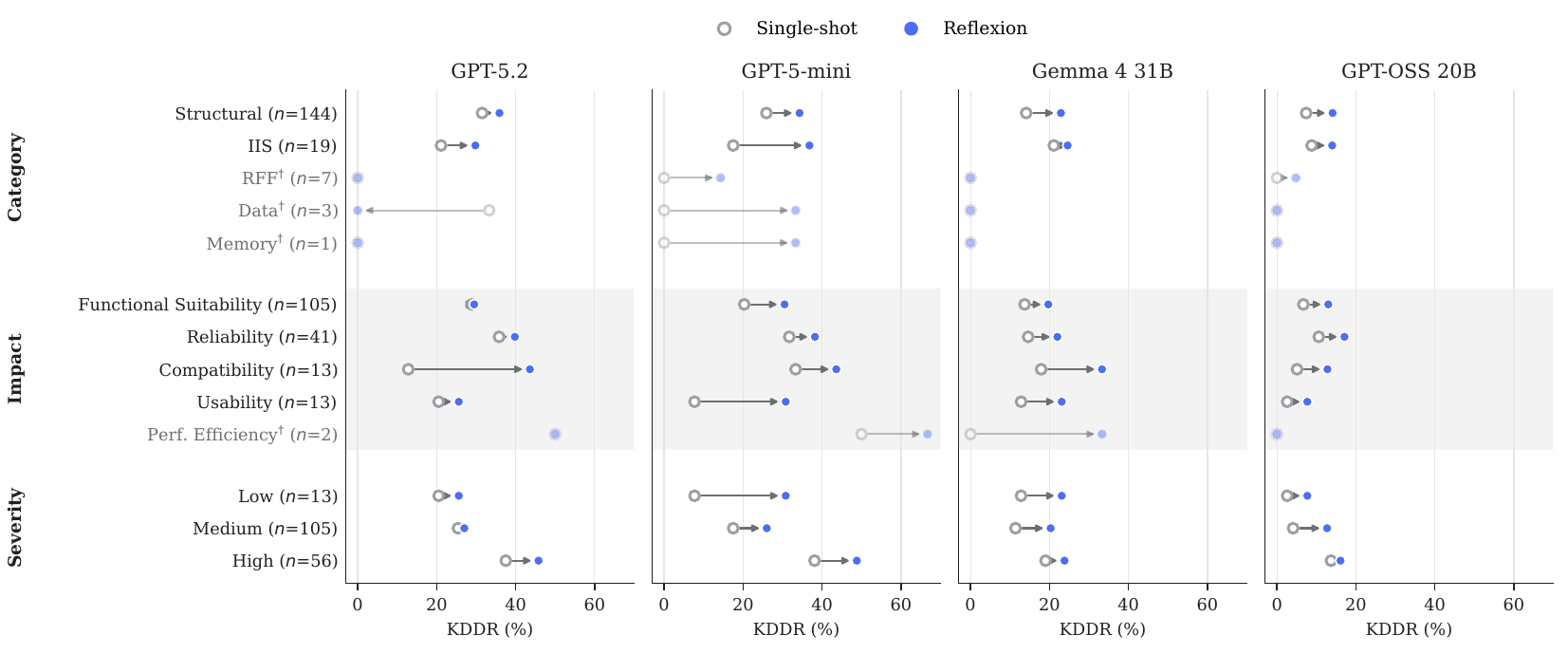}
    \caption{KDDR for every taxonomy group (Table~\ref{tab:kddr_taxonomy}, Appendix~\ref{app:robustness}), one panel per backbone. Hollow gray markers are single-shot, filled blue markers are Reflexion, and arrows point from single-shot to Reflexion. Each value averages the three judges under the balanced prompt. Faded rows marked $^{\dagger}$ have fewer than 10 instances and are shown for completeness only. RFF = Requirements, Features and Functionality, IIS = Interface, Integration and System.}
    \label{fig:kddr_taxonomy}
\end{figure*}

\subsection{Does \texttt{CR-Bench} Reproduce Known Patterns?}
\label{sec:perf}
Both patterns from Section~\ref{sec:experiments} hold. First, Reflexion improves KDDR for every backbone under every judge (Table~\ref{tab:paired_kddr}), significantly in 8 of the 12 backbone--judge pairs. GPT-5.2, the strongest single-shot system, gains 2.9--5.7 points without reaching significance. Second, both GPT-5 backbones beat both open-weight backbones under each strategy (Figure~\ref{fig:pareto}). Gemini Code Assist detects only $8.1 \pm 0.6\%$ of known defects, below every system except single-shot GPT-OSS 20B. Even the best systems detect only about one-third of known defects (KDDR $\approx 33\%$), so \texttt{CR-Bench} is far from saturated.

\subsection{Defect Detection versus Developer Utility}
Reflexion improves both for every backbone, so self-reflection does not trade usefulness for detection. Only GPT-5.2 and GPT-5-mini Reflexion are Pareto-optimal: GPT-5-mini matches GPT-5.2 on KDDR ($33.7$ vs.\ $33.0\%$) but trails it by 17 points in Usefulness Rate. Gemini Code Assist shows the opposite profile: its comments are often useful ($78.0\%$) but rarely hit the known defect.

\subsection{Robustness to the Evaluator}
Each judge scores the same fixed review comments (Figure~\ref{fig:judge_kddr_usefulness}). The judges agree on direction but not on level: GPT-5.4 gives the highest Usefulness Rate for all nine systems and DeepSeek V4 Flash the lowest, while Claude Sonnet 4.5 gives the lowest KDDR for eight of nine. Paired comparisons within a judge are therefore more reliable than absolute scores. For GPT-5.2, the Reflexion KDDR gain also holds under strict and lenient evaluator prompts (Table~\ref{tab:prompt_robustness}, Appendix~\ref{app:robustness}).

\subsection{Which Defects Do Agents Catch?}
Figure~\ref{fig:kddr_taxonomy} breaks KDDR down by taxonomy group. Four groups have fewer than ten instances, so we draw conclusions only from groups with at least $13$. Severity matters only for the GPT-5 backbones: with Reflexion, GPT-5.2 and GPT-5-mini detect $45.8\%$ and $48.8\%$ of high-severity defects against $27.0\%$ and $26.0\%$ of medium ones, while Gemma 4 31B and GPT-OSS 20B show no clear difference. Reflexion improves KDDR in every well-populated group for every backbone, most for GPT-5-mini; the largest single gain is GPT-5.2 on Compatibility defects ($12.8\%$ to $43.6\%$). Table~\ref{tab:kddr_taxonomy} and Figure~\ref{fig:kddr_severity} give the full breakdown.

\subsection{Can an LLM Judge Replace Human Annotators?}
\label{sec:judge_validation}
Three expert developers independently labeled 60 reviews from the GPT-5.2 Reflexion agent. Claude Sonnet 4.5 matches the human majority on $71.7\%$ of them, while pairwise agreement among the experts themselves ranges from 0.48 to 0.75. Under the Alternative Annotator Test~\citep{alt_test}, its winning rate is $67\%$ at $\epsilon=0.3$ and $100\%$ at $\epsilon=0.5$, so once the cost of expert annotation is counted ($\epsilon \geq 0.3$), it is a statistically justified substitute for an individual expert annotator. Appendix~\ref{sec:human_annotation} reports the full study.

\section{Conclusion}
\label{sec:conclusion}

We introduced \texttt{CR-Bench}, a defect-centric $(PR, Review)$ benchmark of 584 instances (174 verified) labeled by root cause, impact, and severity, which scores review agents on known-defect detection and usefulness.

\textbf{Implications for practice.} (1) The best systems catch about one-third of known defects, so agents should assist human reviewers, not gate merges. (2) Self-reflection raised both KDDR and Usefulness Rate for every backbone. (3) Usefulness alone misleads: Gemini Code Assist's comments are useful $78.0\%$ of the time, yet it detects $8.1\%$ of known defects, so test agents against known defects before deployment. (4) LLM judges differ in calibration, so compare agents under the same judge.

Future work will add more languages and agent designs and use \texttt{CR-Bench} for post-training.

% Marks the last line of main content. scripts/eacl_format_check.sh reads this label's
% page from the .aux to enforce the page limit; Limitations and References sit after it
% and do not count. Keep it as the final line of the last main-content section.
\label{sec:endmain}

% Required by ARR/EACL; does not count toward the 8-page content limit.
\section*{Limitations}

\paragraph{Scope of the data.} \texttt{CR-Bench} inherits the scope of SWE-Bench: all instances come from open-source Python repositories, and the 174 \texttt{CR-Bench-Verified} tasks span only 11 of them, with Django alone contributing 63. Results may not transfer to other languages, to proprietary or monorepo codebases, or to review practices outside these projects. The reported agent evaluation uses only the verified split; broader conclusions require evaluating the full corpus and additional ecosystems.

\paragraph{One known defect per instance.} Each instance supplies a single known defect, so KDDR measures whether an agent finds that defect, not whether it finds every defect in a PR. Other correct defect reports are counted as Valid Suggestions rather than Bug Hits and contribute to the Usefulness Rate instead.

\paragraph{LLM-generated references and labels.} Claude Opus 4.5 filters instances for detectability, writes the reference review comments, and assigns the taxonomy labels. Errors in these steps propagate to KDDR and to the per-group results, and the detectability filter may exclude defects that a human reviewer could catch.

\paragraph{Evaluator dependence.} Absolute scores depend on judge calibration: the three judges agree on the direction of differences but not on their magnitude. Our primary human-validated judge, Claude Sonnet 4.5, belongs to the same model family as the model that generated the references, which may favor phrasing close to those references. The human study covers 60 comments from a single system (GPT-5.2 Reflexion) and validates only that judge. We mitigate these effects by reporting three judges, prompt variants, and paired within-judge comparisons.

\paragraph{Contamination.} The repositories, issues, and fixes behind SWE-Bench are public and may appear in the training data of the evaluated models, which could inflate KDDR.

\paragraph{Agents evaluated.} We evaluate two reproducible harnesses, neither of which executes code or uses tools, and one commercial agent whose model and harness are undisclosed and may change over time. We do not report cost or latency, so the extra model calls that Reflexion requires are not weighed against its gains.

\section*{Ethical Considerations}

\paragraph{Data.} \texttt{CR-Bench} is derived from SWE-Bench and SWE-Bench Verified, which are built from public open-source repositories. It adds pull request identifiers, LLM-generated reference comments, and taxonomy labels, and it contains no private or proprietary code.

\paragraph{Human annotation.} Two in-house software engineers and one technical director labeled 60 review comments as Bug Hit, Valid Suggestion, or Noise as part of their regular duties, and they agreed to the use of their labels. The task involved only public code and model-generated review text.

\paragraph{Responsible use.} The best evaluated systems detect about one-third of known defects. Teams that rely on review agents as the only safeguard may therefore merge defective code. We recommend using such agents to assist human reviewers rather than to replace them, and testing them against known defects before deployment.

% Acknowledgments go here, unnumbered (\section*{Acknowledgments}), in the final version only.


\bibliography{custom}

\appendix
% Full-width contents list covering only the appendix sections that follow.
\clearpage
\twocolumn[{%
\section*{Appendix}
\startcontents[appendix]
\printcontents[appendix]{}{1}{\setcounter{tocdepth}{2}}
\vspace{1em}
}]
% The contents list gets a page to itself; Appendix A starts on the next page.
\clearpage

\section{Prompts}
\label{app:prompts}

\subsection{Dataset Creation Prompts}
\label{app:detectability_cr_bench}

\subsubsection{Detectability Filter Prompt}
\label{prompt:detectable_bug}

\begin{promptbox}
(*@\promptrole{System Prompt}@*)
You are a software engineering expert analyzing code changes.
Your task is to determine if a change is of the kind that might have been prevented by reviewing the PR/commit that made that change more carefully or NOT.
Name these PREVENTABLE and UNPREVENTABLE.
PREVENTABLE: Code changes that fix incorrect behavior, errors, crashes, or unexpected results introduced by a previous code change (PR/commit).
UNPREVENTABLE: Everything else including but not limited to: New features, refactoring, documentation, performance improvements, or enhancements.

Respond with JSON only:
{ "category": "PREVENTABLE" or "UNPREVENTABLE", "confidence": 0.0 to 1.0, "reasoning": "brief explanation" }

(*@\promptrole{User Prompt}@*)
Problem Statement:
{problem_statement}

Patch:
{patch}

Is this a bug fix that might have been introduced by a previous code change (PR/commit)?
\end{promptbox}

\subsubsection{Bug Description Generation Prompt}
\label{prompt:issue_description_paraphrase}

\begin{promptbox}
(*@\promptrole{System Prompt}@*)
You are a software engineering expert writing technical bug descriptions.

Given a problem statement and the code fix patch, write a concise technical description of the bug.

**Format Requirements:**
- Start with a clear statement of what's wrong
- Include specific code snippets showing the problematic pattern
- Explain why it's a bug (runtime impact, incorrect behavior)
- Keep it under 100 words
- Use technical language
- Focus on the root cause, not symptoms

Respond with ONLY the bug description (no JSON, no extra formatting).

(*@\promptrole{User Prompt}@*)
Bug Description Task

## Problem Statement
{problem_statement}

## Code Fix
File: {file_path}
Lines: {line_start}-{line_end}

```diff
{patch}
```

Write a concise technical description of this bug.
\end{promptbox}

\subsubsection{Bug Taxonomy Tagging Prompt}
\label{prompt:cateogrize_bugs}

\begin{promptbox}
(*@\promptrole{System Prompt}@*)
You are a software engineering expert analyzing bug reports to classify them according to a predefined bug taxonomy used in empirical software engineering research.

Analyze the bug from the issue description (problem statement) and the code fix patch, and provide a classification using the following fields.

---

### 1. **category**

Choose EXACTLY ONE of the following bug types:

#### **1. Requirements, Features, and Functionality Bugs**

**Source:** Ambiguous, incomplete, misunderstood, or changing requirements/specifications.
**Types:**

* **Requirements bugs:** Missing, ambiguous, or inconsistent requirements.
* **Feature bugs:** Wrong, missing, or extra features, including undesired enhancements.
  **Detection/Remedy:** Formal specification languages, functional testing (transaction flow, domain, logic, state-based testing).

#### **2. Structural Bugs**

**Source:** Code control flow, sequence, and logical structure.
**Types:**

* **Control & sequence bugs:** Missing paths, unreachable code, improper loop nesting, misused GOTO/labels.
* **Logic bugs:** Incorrect logical expressions and Boolean errors.
* **Processing bugs:** Arithmetic, algorithm, type conversion, overflow errors.
* **Initialization bugs:** Uninitialized variables or improper first-loop values.
* **Data-flow anomalies:** Using uninitialized or stale data or failing to store or modify data correctly.
  **Detection/Remedy:** Structural/path testing, unit testing, domain testing.

#### **3. Data Bugs**

**Source:** Specification, format, initial values, and usage of data objects.
**Types:**

* **Dynamic data bugs:** Temporary data issues, shared memory residues.
* **Static data bugs:** Fixed content, parameters, and preprocessing errors.
* **Role-based bugs:** Errors in data used as control, parameter, or information.
* **Attributes, structure, content bugs:** Incorrect semantics, memory allocation, or representation.
  **Detection/Remedy:** Validate all data declarations, centralize shared resources, defensive coding, compile-time checks.

#### **4. Coding Bugs**

**Source:** Programming errors, including wild-card or arbitrary mistakes.
**Types:**

* **Syntax errors:** Usually caught by compilers.
* **Documentation bugs:** Misleading comments or manuals.
  **Detection/Remedy:** Proper code review, accurate documentation, automated syntax checking.

#### **5. Interface, Integration, and System Bugs**

**Source:** Interactions between components, systems, or hardware.
**Types:**

* **External interface bugs:** Wrong device or protocol handling.
* **Internal interface bugs:** Improper module communication.
* **Hardware bugs:** Misunderstood device behavior, I/O errors.
* **OS bugs:** Misuse or assumption of OS services.
* **Software architecture bugs:** Emergent behavior due to module interactions or stress.
* **Integration bugs:** Incompatibilities between integrated modules.
* **System bugs:** Complex interaction of multiple components, rare but expensive.
  **Detection/Remedy:** Design for modularity, stress testing, integration testing, specialist interfaces.

#### **6. Test and Test Design Bugs**

**Source:** Errors in testing processes or test code.
**Types:**

* **Test execution bugs:** Incorrect test implementation.
* **Test design bugs:** Flawed test scenarios or criteria.
  **Detection/Remedy:** Test debugging, automation of test execution and design, test quality assurance.

#### **7. Memory-related Bugs**

**Source:** Improper memory handling.
**Types:**

* **Buffer overflow:** Access beyond allocated memory.
* **Stack smashing:** Overwriting function return addresses.
* **Memory leak:** Lost pointers to allocated memory.
* **Uninitialized read:** Accessing memory before initialization.
* **Double free:** Freeing memory twice.
  **Detection/Remedy:** Careful memory management, automated memory checking tools.

#### **8. Concurrent Bugs**

**Source:** Multi-threading or multi-process synchronization issues.
**Types:**

* **Data race bugs:** Conflicting access to shared memory.
* **Atomicity bugs:** Interrupted sequences of operations.
* **Deadlock:** Processes waiting indefinitely for shared resources.
  **Detection/Remedy:** Proper synchronization, lock ordering, deadlock detection mechanisms.

---

### 2. **severity**

Classify the runtime impact of the bug:

* **Critical:** System crash, severe malfunction, data loss, or serious security risk.
* **High:** Core functionality broken or system unusable in common scenarios.
* **Medium:** Partial malfunction, degraded behavior, or non-critical failures.
* **Low:** Minor issues, cosmetic problems, edge cases, or test-only failures.

---

### 3. **impact**

Choose EXACTLY ONE aspect of the software that the bug directly compromises:

#### **1. Functional Suitability**

* **Description:** Bugs that cause incorrect, missing, or inappropriate behavior relative to requirements.
* **Examples:** Wrong outputs, missing features, incorrect calculations.

#### **2. Performance Efficiency**

* **Description:** Bugs that degrade system performance or resource utilization.
* **Examples:** Slow response, high CPU/memory usage, poor throughput.

#### **3. Compatibility**

* **Description:** Bugs that interfere with interoperability or co-existence with other systems, devices, or software.
* **Examples:** Integration failures, protocol mismatches, API incompatibilities.

#### **4. Usability**

* **Description:** Bugs that make the system confusing, hard to use, or increase user errors.
* **Examples:** Misleading error messages, poor workflow, inaccessible UI elements.

#### **5. Reliability**

* **Description:** Bugs that cause system instability, crashes, incorrect state, or inability to recover from failures.
* **Examples:** Crashes, deadlocks, memory leaks, data corruption.

#### **6. Security**

* **Description:** Bugs that expose the system to threats, unauthorized access, or data compromise.
* **Examples:** Buffer overflows, improper access control, injection vulnerabilities.

#### **7. Maintainability**

* **Description:** Bugs that make code harder to understand, modify, or extend.
* **Examples:** Poorly structured code, missing documentation, complex interdependencies.

#### **8. Portability**

* **Description:** Bugs that prevent software from being installed, executed, or transferred across different environments.
* **Examples:** OS-specific assumptions, hardware incompatibilities, environment-specific failures.

---

### 4. **reasoning**

Concisely explain the reasoning behind the **category**, **impact**, and **severity** of the bug.

---

### Output format

Respond ONLY with valid JSON in the following format:

```json
{
  "category": "...",
  "severity": "...",
  "impact": "...",
  "reasoning": "..."
}
```

Do not include explanations, markdown, or any additional text outside the JSON.

(*@\promptrole{User Prompt}@*)
# Bug Classification Task

## Problem Statement
{problem_statement}

## Code Fix
File: {file_path}
Lines: {line_start}-{line_end}

```diff
{correct_fix}
```

Analyze this bug and provide classification as JSON.
\end{promptbox}

\subsection{Agent Prompts}
\label{app:cr_agents}

\subsubsection{Single-Shot Agent Prompt}
\label{prompt:single_shot_agent}

\begin{promptbox}
(*@\promptrole{System Prompt}@*)
You are an expert code reviewer analyzing pull requests for potential bugs and issues.

Your task is to review the provided code changes and identify:
1. Potential bugs or errors
2. Logic issues
3. Edge cases not handled
4. Security vulnerabilities
5. Performance problems

For each issue found, provide:
- file: The file path
- line: The line number (approximate)
- comment: Clear description of the issue
- severity: "high", "medium", or "low"

Output your review as a JSON array of comments. Example:
[
 {
   "file": "src/module.py",
   "line": 42,
   "comment": "Potential null pointer dereference. Variable 'user' may be None.",
   "severity": "high"
 }
]

If no issues found, return an empty array: []

Focus on actual bugs, not style or formatting issues. Do not be verbose, just provide the review comments in concise manner that descirbes the issue completely.

(*@\promptrole{User Prompt}@*)
# Pull Request Review

**Repository:** {{ repo }}
**PR Number:** #{{ pr_number }}
**Title:** {{ title }}
**Description:** {{ description }}

## Diff
```diff
{{ diff }}
```

Please review the above changes and identify any potential bugs or issues.
\end{promptbox}

% The downloaded experiment source truncates the reflection user prompt after
% "Look for bugs you"; the continuation below is reconstructed from its system
% prompt and output contract rather than claimed as a verbatim recovery.
\subsubsection{Reflexion Agent Prompt}
\label{prompt:reflexion_agent}

\begin{promptbox}
(*@\promptrole{System Prompt}@*)
You are an expert code reviewer specializing in bug detection and security analysis.

Review the provided code changes and identify ONLY actual bugs and issues, NOT style or formatting concerns.

## Focus Areas:
1. **Bugs**: Logic errors, incorrect implementations, type mismatches
2. **Edge Cases**: Unhandled null/undefined, empty arrays, boundary conditions
3. **Security**: SQL injection, XSS, authentication bypasses, data leaks
4. **Concurrency**: Race conditions, deadlocks, thread safety issues
5. **Performance**: Memory leaks, infinite loops, inefficient algorithms

## Output Format:
Return a JSON array. Each issue must include:
- file: Exact file path from the diff
- line: Line number where issue occurs
- comment: Specific, actionable description of the bug (not generic warnings)
- severity: "high" (crashes/security), "medium" (data corruption/incorrect behavior), "low" (minor issues)

## Quality Standards:
- Be SPECIFIC: Point to exact problematic code, not general concerns
- Be CERTAIN: Only report issues you're confident are real bugs
- Explain WHY: Include the consequence of the bug
- Suggest FIX: Briefly mention how to resolve it

Return [] if no bugs found.

Example:
[
  {
    "file": "src/auth.py",
    "line": 23,
    "comment": "Missing null check before accessing 'user.email'. Will throw AttributeError if user is None. Add: if user is None: return error",
    "severity": "high"
  }
]

(*@\promptrole{User Prompt}@*)
# Pull Request to Review

**Repository:** {{ repo }}
**PR #{{ pr_number }}:** {{ title }}

{{ description }}

## Code Changes
```diff
{{ diff }}
```

Analyze the diff above and identify concrete bugs. Focus on what could actually break or cause incorrect behavior.

(*@\promptrole{Reflection System Prompt}@*)
You are performing iterative self-improvement using the Reflexion framework.

Your goal: Produce a MORE COMPLETE and ACCURATE review by finding bugs you missed.

## Reflexion Strategy:

### PRIMARY FOCUS: Find Missed Bugs (False Negatives)
Your previous review likely MISSED important bugs. Search for:
- Null/undefined dereferences you overlooked
- Logic errors in conditionals or loops
- Off-by-one errors, incorrect operators
- Unhandled exceptions or error cases
- Security vulnerabilities (injection, auth bypass)
- Resource leaks, race conditions, memory issues

### SECONDARY: Remove Only OBVIOUS False Positives


**When in doubt, KEEP the issue.** Err on the side of caution.

### TERTIARY: Improve Existing Issues
- Make vague comments more specific
- Adjust severity if clearly wrong
- Add fix suggestions

## Output Requirements:
- Return the FULL JSON array with ALL valid issues
- Prioritize ADDING missed bugs over removing uncertain ones
- Each comment must point to specific problematic code
- Be thorough - it's better to report a potential bug than miss a real one

Return [] only if you're absolutely certain no bugs exist.

(*@\promptrole{Reflection User Prompt}@*)
# Reflexion - Iteration {{ iteration }}: Find What You Missed

**Repository:** {{ repo }} | **PR:** #{{ pr_number }}

## Code to Review
```diff
{{ diff }}
```

---

## Your Previous Review (Iteration {{ prev_iteration }})
You found {{ num_comments }} issue(s):
```json
{{ previous_comments }}
```

---

## Reflexion Task: FIND MISSED BUGS

### PRIMARY: What Bugs Did You Miss?

Carefully re-examine the diff. Look for bugs you may have overlooked in the previous review, including failures that emerge from interactions across changed lines, invalid assumptions, missing validation, incorrect state transitions, and changed call sites. Do not duplicate issues already covered by the previous review.

## Output Your Complete Refined Review

Return a JSON array with:
1. All valid issues from previous review (unless clearly false positive)
2. NEW bugs you discovered in this reflection
3. Improved comments and severity levels

Format:
```json
[
  {
    "file": "path/to/file.py",
    "line": 42,
    "comment": "Specific bug: what breaks, why, and how to fix",
    "severity": "high|medium|low"
  }
]
```

Return [] only if absolutely no bugs exist.
\end{promptbox}

\subsection{Evaluation Prompts}
\label{app:cr_verifier}

\subsubsection{LLM-as-a-Judge Prompt -- Balanced}
\label{prompt:cr_verifier}

\begin{promptbox}
(*@\promptrole{System Prompt}@*)
You are a careful code review evaluation assistant.
You classify review comments against a known bug.
Respond with valid JSON only and avoid extra prose.

(*@\promptrole{User Prompt}@*)
You are evaluating code review comments against a known bug.

## Known Bug Description:
{bug_description}

## Files Changed in Fix:
{patch_files}

## Review Comment to Evaluate:
File: {file}
Line: {line}
Comment: {comment}

## Task:
Classify this review comment into ONE of these categories:

1. **BUG_HIT**: The comment identifies or relates to the same issue described in the bug description
2. **VALID_SUGGESTION**: The comment makes a valid point (style, performance, maintainability, edge case, etc.) but is NOT about the known bug
3. **NOISE**: The comment is incorrect, irrelevant, or not actionable, only select this if really feel annoying or not helpful else classify as valid.

Respond with ONLY a JSON object:
{"classification": "BUG_HIT|VALID_SUGGESTION|NOISE", "reason": "brief explanation"}
\end{promptbox}

\subsubsection{LLM-as-a-Judge Prompt -- Strict}
\label{prompt:cr_verifier_strict}

\begin{promptbox}
(*@\promptrole{System Prompt}@*)
You are a code review evaluation assistant.
You classify review comments against a known bug.
Return valid JSON only.

(*@\promptrole{User Prompt}@*)
You are a code review classifier.

Known bug:
{bug_description}

Files changed in fix:
{patch_files}

Review comment:
File: {file}
Line: {line}
Comment: {comment}

Decision rubric:

BUG_HIT:

* Mentions the same defect
* Identifies the same edge case
* Describes the same incorrect behavior
* Points to the same logical flaw

VALID_SUGGESTION:

* Correct and actionable
* Improves maintainability, style, readability, safety, or performance
* Discusses a different issue than the known bug

NOISE:

* Incorrect analysis
* Non-actionable
* Irrelevant
* Pure speculation without evidence

Output ONLY:
{"classification":"BUG_HIT|VALID_SUGGESTION|NOISE","reason":"brief explanation"}
\end{promptbox}

\subsubsection{LLM-as-a-Judge Prompt -- Lenient}
\label{prompt:cr_verifier_lenient}

\begin{promptbox}
(*@\promptrole{System Prompt}@*)
You are a generous but honest code review evaluation assistant.
Give credit to comments that are meaningfully related or clearly useful, but still distinguish true bug hits from general suggestions.
Return valid JSON only.

(*@\promptrole{User Prompt}@*)
You are evaluating whether a code review comment matches a known bug.

Known bug:
{bug_description}

Files changed:
{patch_files}

Review comment:
File: {file}
Line: {line}
Comment: {comment}

First determine:

1. What issue the review comment is describing
2. Whether that issue matches the known bug

Then classify as:

* BUG_HIT
* VALID_SUGGESTION
* NOISE

Return ONLY:
{"classification":"BUG_HIT|VALID_SUGGESTION|NOISE","reason":"brief explanation"}
\end{promptbox}

% Full-width block at the top of a fresh page keeps the heading and the table together;
% a table* float would be pushed to the following page.
\clearpage
\twocolumn[{%
\section{Novelty Positioning Matrix}
\label{app:novelty}

Table~\ref{tab:novelty-matrix} positions \texttt{CR-Bench} against prior code review datasets on the four properties discussed in Section~\ref{sec:related_work}: multiple diff hunks, full PR context, categorization, and a focus on a verifiable defect.

\vspace{0.75em}
\begin{minipage}{\textwidth}
    \centering
    \small
    \setlength{\tabcolsep}{4pt}
    \begin{tabularx}{\textwidth}{@{}>{\raggedright\arraybackslash}p{0.19\textwidth}c*{4}{>{\centering\arraybackslash}X}>{\raggedright\arraybackslash}X@{}}
        \toprule
        \textbf{Method} & \textbf{\# Tasks} & \textbf{Multiple Diffs} & \textbf{Full PR Context} & \textbf{Categorization} & \textbf{Defect-focused} & \textbf{Evaluation} \\
        \midrule
        \citealp{trans-review-data} & $1718$ & \ding{55} & \ding{55} & \ding{55} & \ding{55} & Exact/BLEU \\
        \citealp{t5-codereview} &$ 147,533$ & \ding{55} & \ding{55} & \ding{55} & \ding{55} & Exact/BLEU \\
        \citealp{autotransform} & $~17K$ & \ding{55} & \ding{55} & \ding{55} & \ding{55} & Exact \\
        \citealp{codereviewer} & $54K$ & \ding{55} & \ding{55} & \ding{55} & \ding{55}  & CodeBLEU \\
        \citealp{swr-bench} & $1000$ & \ding{51} & \ding{51} & \ding{51} & \ding{55} & LLM Judge (P, R, F1) \\
        \midrule
        \textbf{CR-Bench (Ours)} & 584 & \ding{51} & \ding{51} & \ding{51} & \ding{51} & LLM Judge (KDDR, U) \\
        \textbf{CR-Bench-Verified (Ours)} & $174$ & \ding{51} & \ding{51} & \ding{51} & \ding{51} & LLM Judge (KDDR, U) \\
        \bottomrule
    \end{tabularx}
    \captionof{table}{A novelty matrix positioning \texttt{CR-Bench} compared to existing code review datasets.}
    \label{tab:novelty-matrix}
\end{minipage}
\vspace{1.5em}
}]

\section{Dataset}
\label{app:dataset}
\label{app:cr_bench_stats}

This appendix reports dataset statistics and an example instance.

\subsection{Statistics}

\begin{figure*}[t]
    \centering
    \includegraphics[width=0.96\linewidth]{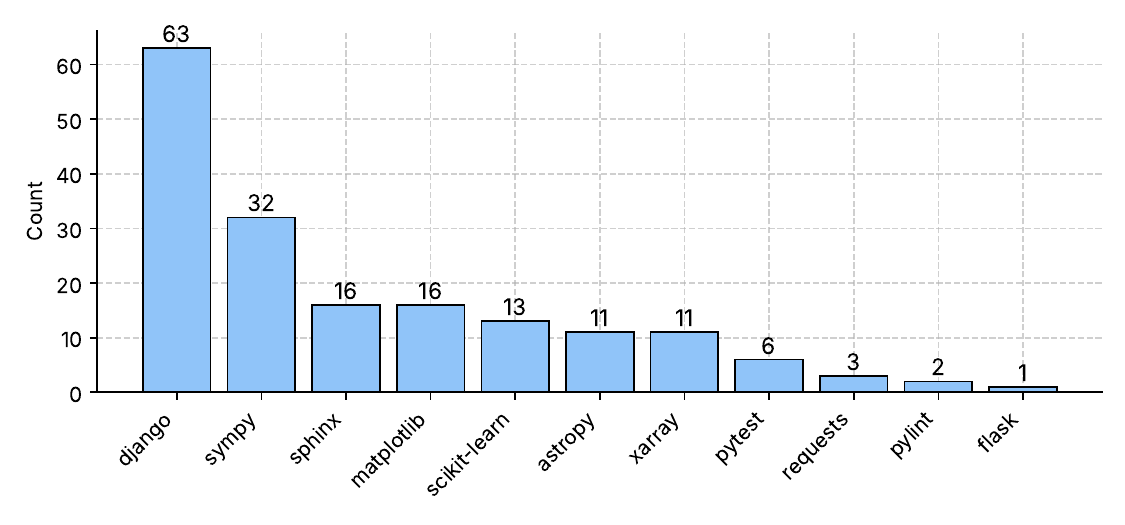}
    \caption{Tasks per source repository in \texttt{CR-Bench-Verified} (174 instances). Repositories are inherited from SWE-Bench Verified. Django is the largest repository with 63 tasks (36.2\%), and Django and SymPy together account for just over half of the benchmark (54.6\%). The remaining nine repositories span documentation tooling (Sphinx), scientific computing (Matplotlib, scikit-learn, Astropy, xarray), testing and linting (pytest, pylint), and HTTP and web libraries (Requests, Flask).}
    \label{fig:histogram}
\end{figure*}

Table~\ref{tab:crbench_stats} reports per-instance means for both splits, and Figure~\ref{fig:histogram} shows how the \texttt{CR-Bench-Verified} tasks are distributed across source repositories.

\begin{table}[!htb]
\centering
\small
\setlength{\tabcolsep}{4pt}
\begin{tabular}{lcc}
\toprule
\textbf{Dataset} & \textbf{Fixed} & \textbf{PR} \\
 & \textbf{lines} & \textbf{comments} \\
\midrule
\texttt{CR-Bench} & 10.28 & 41.03 \\
\texttt{CR-Bench-Verified} & 8.69 & 35.83 \\
\bottomrule
\end{tabular}
\caption{Per-instance means for all 584 \texttt{CR-Bench} instances and the 174-instance verified subset. Fixed lines count lines changed by the known repair. PR comments count comments in the originating pull request.}
\label{tab:crbench_stats}
\end{table}

\subsection{Example Instance}

\begin{figure*}[t]
    \centering
    \begin{subfigure}[t]{0.72\linewidth}
        \centering
        \includegraphics[width=\linewidth]{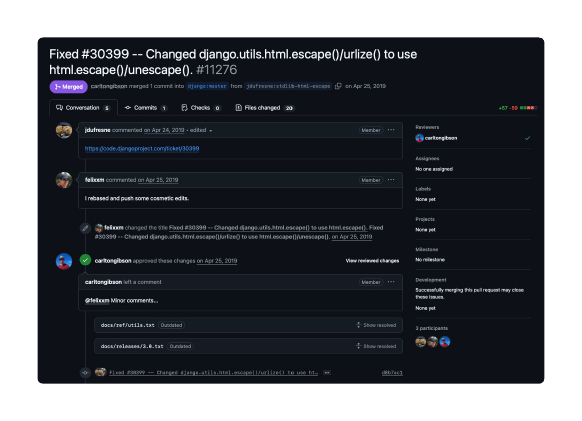}
        \caption{Originating pull request (PR)}
        \label{fig:pr}
    \end{subfigure}

    \vspace{0.5em}
    \begin{subfigure}[t]{0.47\linewidth}
        \centering
        \includegraphics[width=\linewidth]{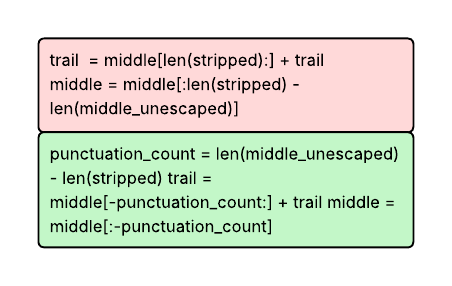}
        \caption{Known repair patch ($r_\delta$)}
        \label{fig:patch}
    \end{subfigure}
    \hfill
    \begin{subfigure}[t]{0.47\linewidth}
        \centering
        \includegraphics[width=\linewidth]{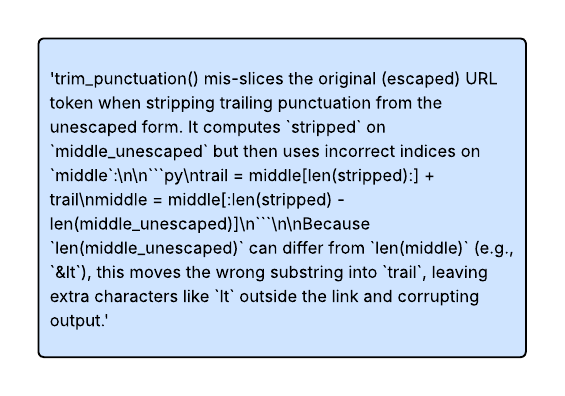}
        \caption{Target review comment ($r_c$), labeled Bug Hit}
        \label{fig:comment}
    \end{subfigure}

    \caption{One \texttt{CR-Bench} instance. The PR context (top) concerns Django's escaped-URL punctuation handling. Its known defect is that \texttt{trim\_punctuation()} slices escaped URLs with indices derived from the unescaped string. The repair $r_\delta$ (bottom left) computes the removed punctuation count before slicing. Red and green mark removed and added code. The target comment $r_c$ (bottom right) identifies the same defect, so its label is Bug Hit.}
    \label{fig:example}
\end{figure*}

Figure~\ref{fig:example} shows one \texttt{CR-Bench} instance: the pull request, the known fix patch, and the reference review comment.

% Two rows keep the PR context and the decisive patch/comment text legible at print size.

% Start on a fresh page so the Dataset section's floats are placed before this one.
\clearpage
\section{Taxonomy}
\label{app:taxonomy}

\begin{figure*}[t]
    \centering
    \includegraphics[width=0.96\linewidth]{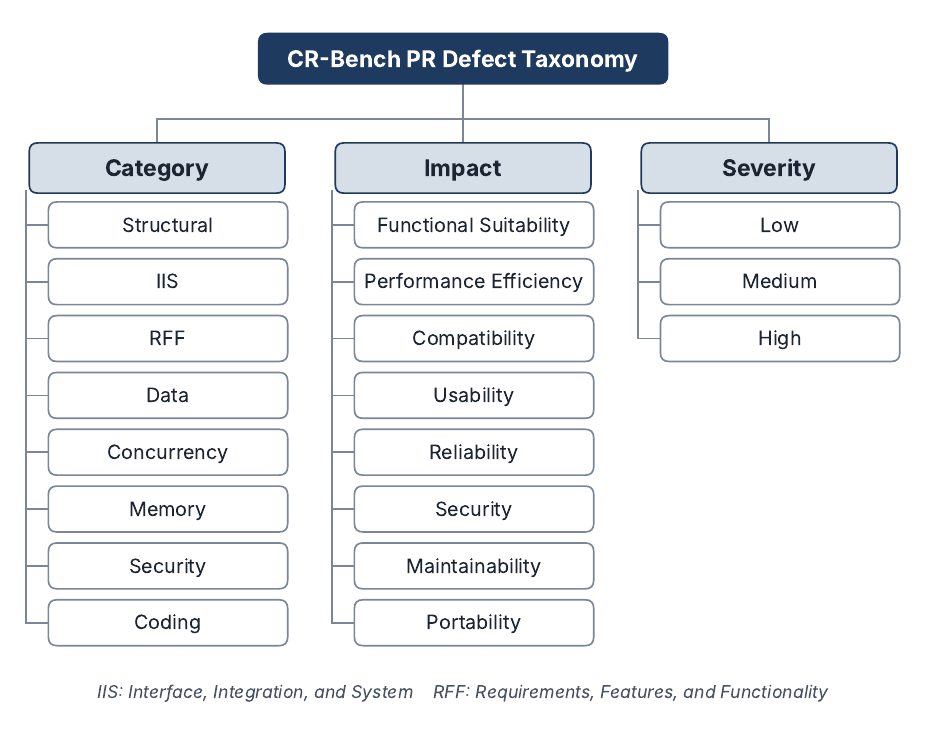}
    \caption{Taxonomy Labels for PR defects including Category, Impact, and Severity.}
    \label{fig:taxonomy}
\end{figure*}

\begin{figure*}[t]
    \centering
    \includegraphics[width=\linewidth]{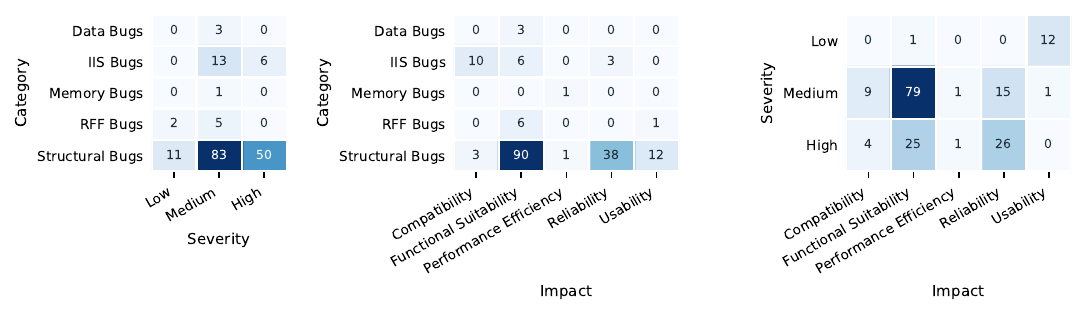}
    \caption{Pairwise counts for the 174 \texttt{CR-Bench-Verified} instances. Panels show category by severity (left), category by impact (center), and severity by impact (right). Only observed taxonomy levels appear. RFF = Requirements, Features and Functionality. IIS = Interface, Integration and System.}
    \label{fig:dist_verified}
\end{figure*}

\begin{figure*}[htbp]
    \centering
    \includegraphics[width=\linewidth]{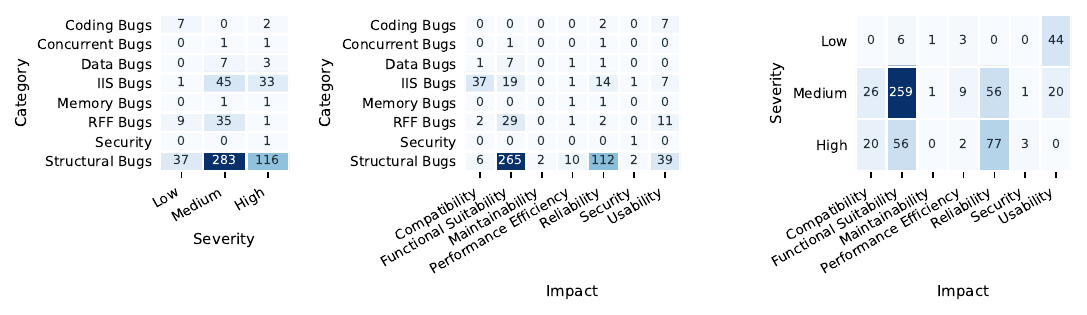}
    \caption{Pairwise cross-tabulations for all 584 \texttt{CR-Bench} instances. The panels show category by severity (left), category by impact (center), and severity by impact (right). Each cell counts instances. Only observed taxonomy levels are shown. RFF = Requirements, Features and Functionality. IIS = Interface, Integration and System.}
    \label{fig:dist_full}
\end{figure*}

Figure~\ref{fig:taxonomy} lists the taxonomy labels assigned to each defect. Figures~\ref{fig:dist_verified} and~\ref{fig:dist_full} show how the labels co-occur in \texttt{CR-Bench-Verified} and \texttt{CR-Bench}.

% Start on a fresh page so earlier sections' floats are placed before this one.
\clearpage
\section{Alternative Annotator Test}
\label{sec:human_annotation}
\label{app:human_alignment}

We use \textbf{Alt-Test}~\citep{alt_test} to evaluate the suitability of an LLM as a replacement for a human code reviewer.

\subsection{Annotation Setup}

To verify the quality of \texttt{CR-Bench}'s LLM-as-a-judge, Claude Sonnet 4.5, we study its alignment with humans. We employ $M=3$ expert developers to independently annotate a subsample of $60$ reviews from Reflexion agent with GPT-5.2 (Table~\ref{tab:human_review_example} shows one annotated instance).

% Full-width float: the example instance contains long unbreakable code lines.
\begin{table*}[t]
\centering
\footnotesize
\renewcommand{\arraystretch}{1.3}
\begin{tabularx}{\linewidth}{|p{0.28\linewidth}|>{\RaggedRight\arraybackslash}X|}
\hline

\textbf{instance\_id} & django\_\_django-12708 \\ \hline
\textbf{Repository} & django/django \\ \hline
\textbf{Bug Origin PR} & 8432 \\ \hline

\textbf{PR Link} & 
\url{https://github.com/django/django/pull/8432} \\ \hline

\textbf{Known Bug Description} & 
Deleting an \texttt{index\_together} could crash when the same columns are also in \texttt{unique\_together} because the schema editor searched for any constraint matching the columns and \texttt{index=True}, which also matches unique constraints. In \texttt{alter\_index\_together()} it called:

\texttt{self.\_delete\_composed\_index(model, fields, \{'index': True\}, self.sql\_delete\_index)}

This can return both the unique and non-unique index constraints, causing \texttt{\_delete\_composed\_index()} to raise \texttt{ValueError} due to multiple matches. The fix constrains the lookup to non-unique indexes (\texttt{\{'index': True, 'unique': False\}}), ensuring only the intended index is deleted.
\\ \hline

\textbf{Known Fix Patch} & 
\parbox[t]{\linewidth}{\ttfamily\tiny
diff --git a/django/db/backends/base/schema.py b/django/db/backends/base/schema.py\\
--- a/django/db/backends/base/schema.py\\
+++ b/django/db/backends/base/schema.py\\
@@ -393,7 +393,12 @@ def alter\_index\_together(self, model,\\
 old\_index\_together, new\_index\_together):\\
 news = \{tuple(fields) for fields in new\_index\_together\}\\
 \# Deleted indexes\\
 for fields in olds.difference(news):\\
 - self.\_delete\_composed\_index(model, fields, \{'index': True\}, self.sql\_delete\_index)\\
 + self.\_delete\_composed\_index(\\
 +     model,\\
 +     fields,\\
 +     \{'index': True, 'unique': False\},\\
 +     self.sql\_delete\_index,\\
 + )\\
}
\\ \hline

\textbf{Files Changed in Fix} & django/db/backends/base/schema.py \\ \hline
\textbf{Review File} & django/contrib/auth/backends.py \\ \hline
\textbf{Review Line} & 57 \\ \hline

\textbf{Review Comment} & 
Potential compatibility risk if this branch is intended to support Python <2.7: set literals and dict comprehensions used throughout this PR (e.g., \texttt{"\%s.\%s" \% (ct, name) for ct, name in perms}) require Python 2.7+/3.x. If any supported runtime is older, this will raise \texttt{SyntaxError} at import time. Fix by keeping \texttt{set(... for ...)} / \texttt{dict(...)}.
\\ \hline
\rowcolor{gray!20}
\textbf{Review Type Label} & \textbf{Noise} \\ \hline

\end{tabularx}
\caption{An example of a human review annotation}
\label{tab:human_review_example}
\end{table*}

\subsection{Formalism}

We formalize code review evaluation as a classification task over a set of annotators, comprising both LLMs and human reviewers. Let $x \in \mathcal{X}$ denote a code review instance and $y \in \mathcal{Y}$ denote a review label. Each annotator $a \sim \mathcal{A}$ induces a conditional distribution $P_a(y \mid x)$.

Given the inherently subjective nature of code review, label assignment is influenced by annotator-specific biases and contextual factors. To capture this variability, we define the aggregated label distribution as:
\begin{equation}
    \hat{P}(y \mid x) = \frac{1}{|\mathcal{A}|} \sum_{a \in \mathcal{A}} P_a(y \mid x).
\end{equation}

We quantify the degree of subjectivity for an instance $x$ as the entropy of the aggregated distribution:
\begin{equation}
\begin{split}
\mathcal{S}(x) &= H(\hat{P}(y \mid x)) \\
&= - \sum_{y \in \mathcal{Y}} \hat{P}(y \mid x)\log \hat{P}(y \mid x).
\end{split}
\end{equation}

\paragraph{Individual Annotator Evaluation.} Given an annotator $a$ producing prediction $y_a(x)$, this approach compares this prediction against the judgments $\{R_i\}_{i=1}^M$ of $M$ independent annotators using parameterized majority voting. The agreement score is defined as ($k$ is a parameter):

\begin{equation}
\text{Agree}(x, a) = 
\begin{cases} 
1 & \text{if } \sum_{i=1}^{M} \mathbb{I}(R_i = y_a) \ge k \\ 
0 & \text{otherwise}.
\end{cases}
\label{eq:agreement}
\end{equation}

This formulation implicitly treats consensus labels as ground truth, thereby collapsing the underlying distribution of valid annotations into a single point estimate and introducing majority bias in subjective settings. However, annotator disagreement is intrinsic to such tasks and reflects a latent distribution rather than noise. To account for this, we adopt the Alternative Annotator Test (Alt-Test)~\citep{alt_test}, which evaluates annotator quality relative to the inter-annotator agreement baseline instead of a fixed consensus label.

The Alt-Test excludes each human annotator and compares LLM alignment with the remaining. It needs two parameters: the cost-benefit penalty ($\epsilon$) and the False Discovery Rate threshold ($q$). $\epsilon$ quantifies the cost-benefit trade-off of using an LLM. A higher $\epsilon$ means a larger associated penalty to using human annotators, maybe because they are more expensive, slow, etc. We evaluate the system across a range of penalties ($\epsilon \in [0.0, 1.0]$) and report winning rate $\omega$, the fraction of annotators an LLM can be better than. We use $q=0.05$, meaning we strictly limit the probability of falsely concluding the LLM as a valid replacement to $5\%$.

\subsection{Results}

\begin{figure*}[t]
    \centering
    % Left Subfigure: Epsilon Sweep
    \begin{subfigure}[b]{0.48\textwidth}
        \centering
        \includegraphics[width=\textwidth]{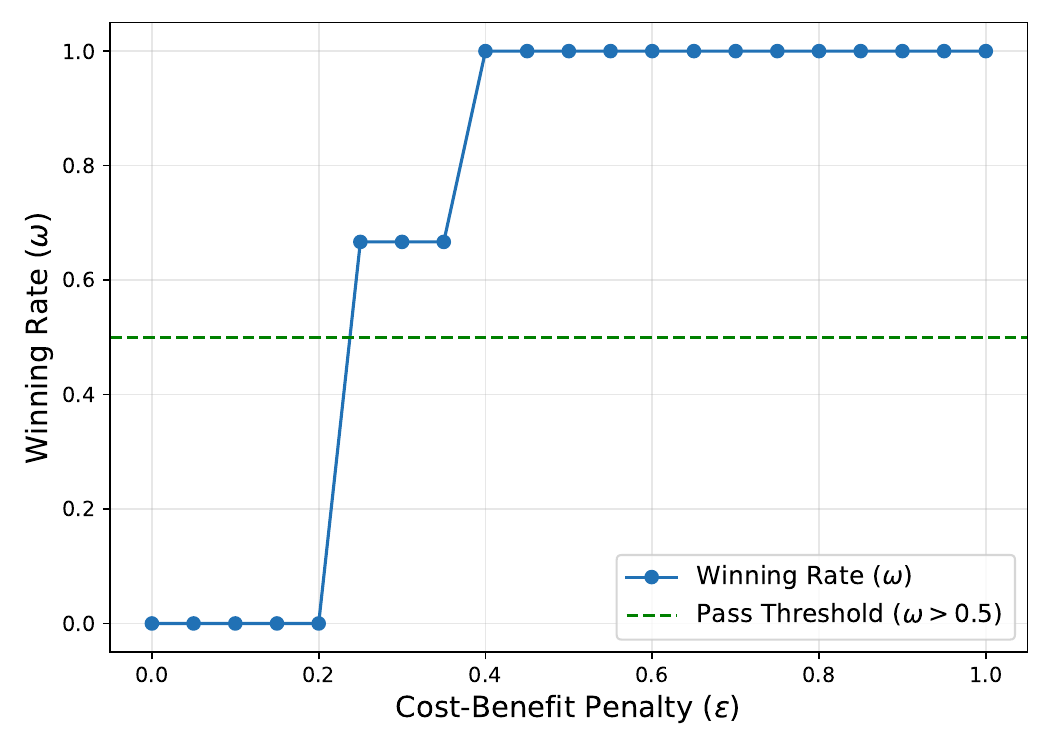}
        \caption{Winning Rate ($\omega$) vs. Cost-Benefit Penalty ($\epsilon$)}
        \label{fig:alt_test_epsilon}
    \end{subfigure}
    \hfill
    % Right Subfigure: Sample Size Sweep
    \begin{subfigure}[b]{0.48\textwidth}
        \centering
        \includegraphics[width=\textwidth]{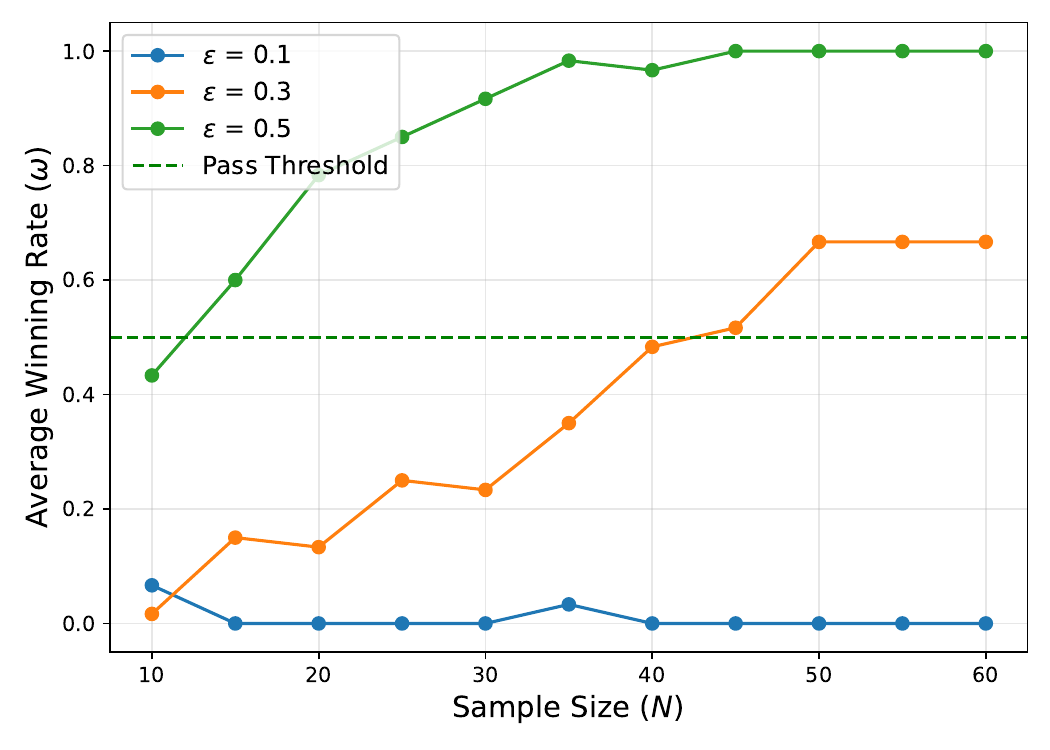}
        \caption{Winning Rate ($\omega$) vs. Sample Size ($N$)}
        \label{fig:alt_test_sample}
    \end{subfigure}

    \caption{\textbf{Alternative Annotator Test (Alt-Test) Results.}
    \textbf{(a)} The LLM's winning rate ($\omega$) across varying cost-benefit penalties ($\epsilon$) ($N=60$). The LLM successfully replaces the human majority ($\omega > 0.5$) at $\epsilon \ge 0.25$.
    \textbf{(b)} The effect of sample size on winning rate, averaged over 20 random iterations. Test reaches statistical significance within the available data.}
    \label{fig:alt_test_results}
\end{figure*}

% Was a wraptable in the single-column COLM layout; a wrapped float does not fit an
% ACL column, so it is now a normal single-column float.
\begin{table}[!htb]
    \centering
    \small
    \begin{tabular}{c|c|c}
        \toprule
        Metric & Setting & Value($\%$) \\
        \midrule
        \multirow{3}{*}{IAA score} 
        & $k=1$ & $71.66\%$ \\
        & $k=2$ & $71.66\%$ \\
        & $k=3$ & $28.33\%$ \\
        \midrule
        \multirow{3}{*}{Winning rate} 
        & $\epsilon=0.1$ & $0\%$ \\
        & $\epsilon=0.3$ & $67\%$ \\
        & $\epsilon=0.5$ & $100\%$ \\
        \bottomrule
    \end{tabular}
    \caption{IAA score and Alt-Test winning rate($\omega$)~\citep{alt_test} for classifying reviews using \texttt{claude-4.5-sonnet}.}
    \label{tab:result_comparison}
\end{table}

As per IAA score in Table~\ref{tab:result_comparison}, model can't make a strict agreement with all the annotators ($k=3$), but is able to maintain a consistency at lesser thresholds. With the $\omega$ analysis using Alt-Test, we further determine that LLM cannot statistically justify replacing humans at low $\epsilon$ ($\omega=0\%$). However, as $\epsilon$ increases to simulate the operational benefits of automation, the model's $\omega$ climbs to 67\% at $\epsilon=0.3$ and reaches $100\%$ at $\epsilon=0.5$.

As per experienced software engineer annotators' feedback, this is a complex and time-taking labeling. So, consistent with the Alt-Test paper, this is $\epsilon \geq 0.2$. Further, $\omega\geq50\%$ translates to LLMs being able to replace humans. We sweep $\omega$ as a function of $\epsilon$ in Figure~\ref{fig:alt_test_epsilon} and show that our judge achieves $67\%$ $\omega$ at $\epsilon=0.3$ and $100\%$ at $\epsilon=0.5$. $\omega=0$ for $\epsilon\leq0.2$ shows that code-review of this standard is a difficult task for even skilled humans to annotate.

Further, as discussed earlier, we only consider $N=60$ out of approx. $1600$ reviews generated from $174$ PRs for this study. In order to understand its sufficiency, we plot a graph of $\omega$ v/s $N$. For each $n\in N$, we randomly sample $n$ instances and calculate $\omega$ for a given $\epsilon$. We do this $20$ times and plot the average in Figure~\ref{fig:alt_test_sample}. Saturation of $\omega$ after $N\geq50~\forall~\epsilon~\in~\{0.1,0.3,0.5\}$ shows that $N=60$ is sufficient for statistical significance.

\subsection{Agreement Analyses}

\begin{figure}[!htb]
    \centering
    \includegraphics[width=\columnwidth]{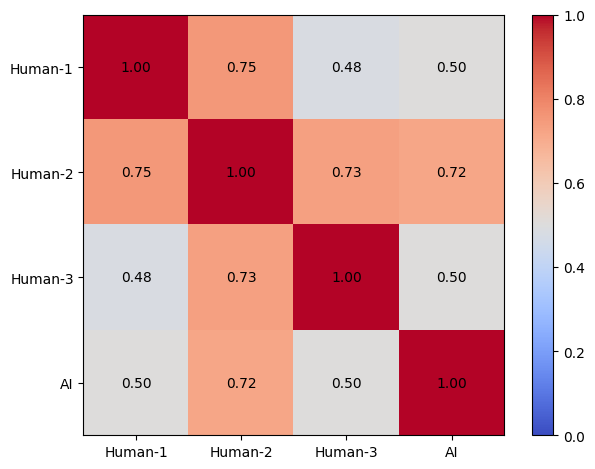}
    \caption{Pairwise raw agreement over 60 review comments labeled as Bug Hit, Valid Suggestion, or Noise by three human experts and Claude Sonnet 4.5. Each cell is the proportion of identical labels. Diagonal cells show self-agreement.}
    \label{fig:IAA}
\end{figure}

\begin{figure}[!htb]
    \centering
    \includegraphics[width=\columnwidth]{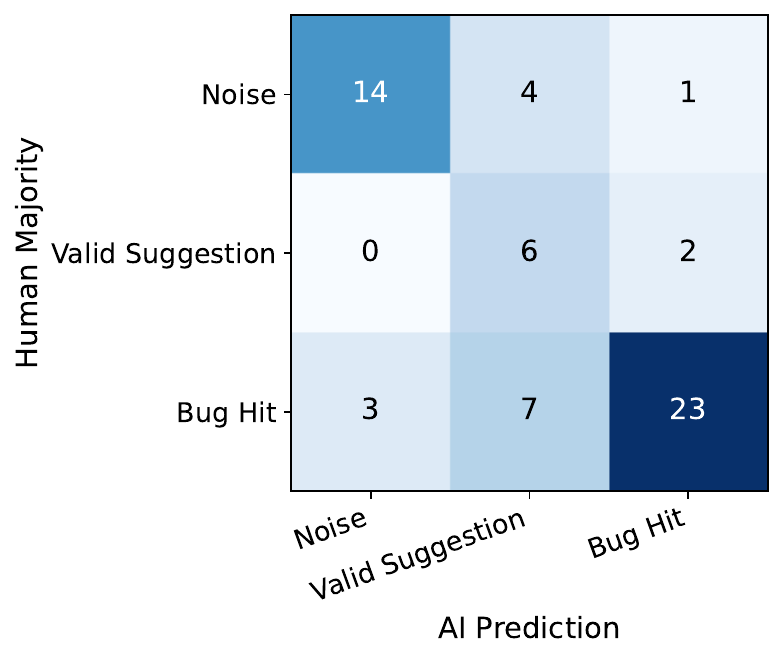}
    \caption{Human majority vs. AI prediction confusion matrix across three categories (Noise, Valid Suggestion, Bug Hit). Each cell reports the number of instances with a given human majority label (rows) and corresponding AI prediction (columns). The strong diagonal for Bug Hit indicates high alignment on objective cases, while off-diagonal mass, particularly from Bug Hit to Valid Suggestion, highlights systematic ambiguity and the AI's tendency to produce softer, non-definitive judgments in borderline instances.}
    \label{fig:confusion_matrix}
\end{figure}

\begin{figure}[!htb]
    \centering
    \includegraphics[width=\columnwidth]{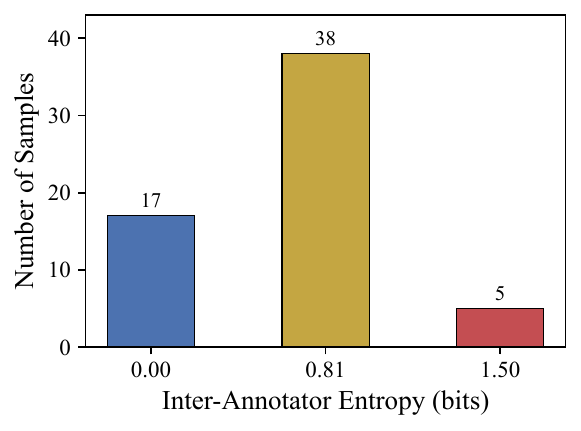}
    \caption{Distribution of per-sample inter-annotator entropy across 60 code review comments, computed over four annotators (three human, one AI). An entropy of 0.00 bits indicates unanimous agreement, 0.81 bits reflects a 3-vs-1 split, and 1.50 bits indicates a three-way split across all label categories. The majority of samples (63\%) 
  exhibit partial disagreement with a single dissenting annotator. }
    \label{fig:iaa_entropy_dist} % was named as the same.
\end{figure}

\begin{figure}[!htb]
    \centering
    \includegraphics[width=\columnwidth]{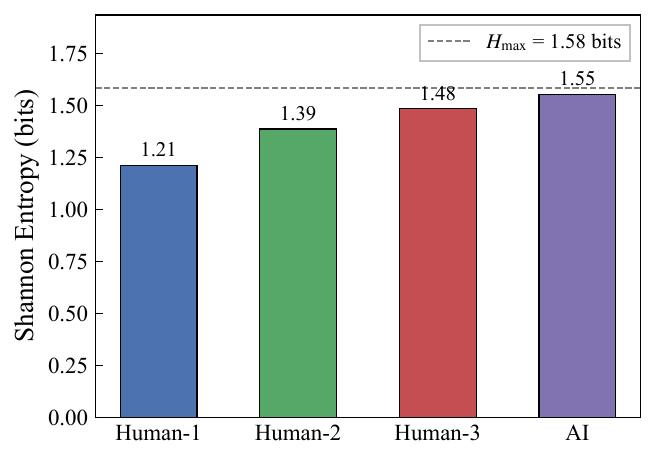}
    \caption{Shannon entropy of the label distribution for each annotator, computed over 60 code review comments across three categories: Bug Hit, Noise, and Valid Suggestion. The dashed line denotes the maximum entropy ($H_{\max}$ = 1.58 bits) for a uniform distribution over three classes. Human-1 exhibits the most concentrated labeling (1.21 bits), while AI produces the most uniform distribution (1.55 bits), suggesting greater diversity in its label assignments. }
    \label{fig:iaa_entropy_label}
\end{figure}

Further analyses also show that classifying review comments as Bug Hit, Valid Suggestion, or Noise is a difficult task even for expert annotators. Pairwise agreement between the expert annotators themselves spans 0.48 to 0.75, and the judge aligns most closely with the most central annotator (Figure~\ref{fig:IAA}), which suggests it approximates consensus behavior rather than any one reviewer. The confusion matrix against the human majority (Figure~\ref{fig:confusion_matrix}) shows strong agreement on Bug Hits, with most residual error being Bug Hits softened into Valid Suggestions. Figure~\ref{fig:iaa_entropy_dist},~\ref{fig:iaa_entropy_label} show inter and intra-annotator entropy of AI falls in similar range with majority humans.

Pairwise agreement among the human annotators ranges from 0.48 to 0.75. Human-2 has the highest agreement with the other annotators. Claude Sonnet 4.5 has its highest human agreement with Human-2 (0.72). This pattern is consistent with consensus-oriented labeling.

% Start on a fresh page so earlier sections' floats are placed before this one.
\clearpage
\section{Results}
\label{app:results}
\label{app:robustness}

\subsection{Evaluator Robustness and Uncertainty}

All results below use clean judge runs with zero fallback classifications. To separate review-system performance from evaluator choice, each judge independently classifies the same fixed review comments. Table~\ref{tab:judge_exact} reports every review-system--judge combination, while Table~\ref{tab:paired_all_metrics} directly compares Reflexion with single-shot for matched backbones. Confidence intervals resample benchmark instances. Paired tests align strategies by \texttt{instance\_id}.

% Detailed judge-level visualization supports the compact aggregate results in the main paper.
\begin{figure*}[t]
    \centering
    \includegraphics[width=0.96\linewidth]{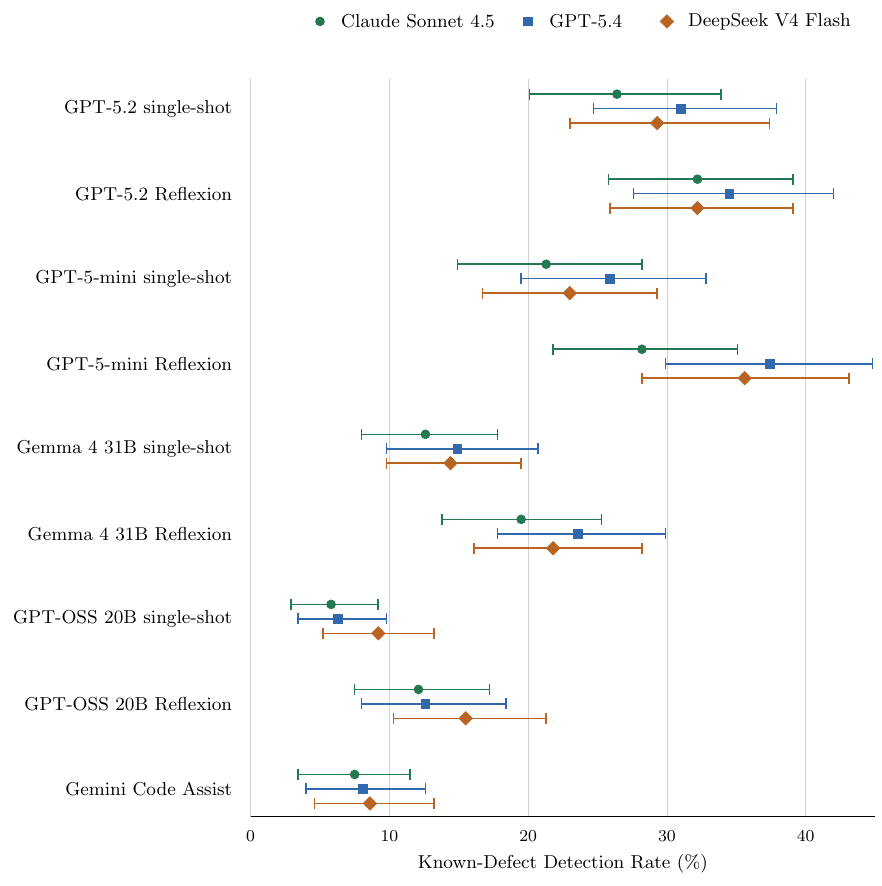}
    \caption{Judge-specific Known-Defect Detection Rate (KDDR) for the same fixed review comments on all 174 verified instances under the balanced prompt. Points are full-sample estimates. Bars are 95\% percentile confidence intervals from 1,000 instance-level bootstrap resamples. Marker shape identifies the judge independently of color.}
    \label{fig:kddr_judge_forest}
\end{figure*}

% Full-width float: these metric tables are far wider than one ACL column.
\begin{table*}[t]
\centering
\small
\begin{tabular}{lcc}
\toprule
\textbf{System} & \textbf{KDDR} & \textbf{Usefulness Rate} \\
\midrule
GPT-5.2 single-shot & 28.9 $\pm$ 2.3 & 86.2 $\pm$ 11.3 \\
GPT-5.2 Reflexion & 33.0 $\pm$ 1.3 & 89.8 $\pm$ 9.5 \\
\midrule
GPT-5-mini single-shot & 23.4 $\pm$ 2.3 & 69.0 $\pm$ 29.9 \\
GPT-5-mini Reflexion & 33.7 $\pm$ 4.9 & 72.6 $\pm$ 28.4 \\
\midrule
Gemma 4 31B single-shot & 14.0 $\pm$ 1.2 & 64.5 $\pm$ 27.8 \\
Gemma 4 31B Reflexion & 21.6 $\pm$ 2.1 & 68.6 $\pm$ 27.3 \\
\midrule
GPT-OSS 20B single-shot & 7.1 $\pm$ 1.8 & 53.3 $\pm$ 27.9 \\
GPT-OSS 20B Reflexion & 13.4 $\pm$ 1.8 & 61.2 $\pm$ 26.6 \\
\midrule
Gemini Code Assist & 8.1 $\pm$ 0.6 & 78.0 $\pm$ 20.0 \\
\bottomrule
\end{tabular}
\caption{Performance on \texttt{CR-Bench-Verified}. Each cell is the mean $\pm$ sample standard deviation over the three judges (GPT-5.4, Claude Sonnet 4.5, DeepSeek V4 Flash, $n=3$) under the balanced prompt. The deviation measures spread across evaluators, not sampling error; per-judge estimates with instance-bootstrap confidence intervals are in Table~\ref{tab:judge_exact}. Percentage metrics omit the \% symbol.}
\label{tab:results}
\end{table*}

\begin{table*}[t]
\centering
\scriptsize
\resizebox{\textwidth}{!}{%
\begin{tabular}{@{}llcc@{}}
\toprule
Review system & Evaluation judge & KDDR (\%) [95\% CI] & Usefulness Rate (\%) [95\% CI] \\
\midrule
GPT-5.2 single-shot & Claude Sonnet 4.5 & 26.4 [20.1,33.9] & 81.7 [77.9,85.5] \\
 & GPT-5.4 & 31.0 [24.7,37.9] & 99.1 [98.6,99.5] \\
 & DeepSeek V4 Flash & 29.3 [23.0,37.4] & 77.8 [75.5,80.4] \\
\addlinespace
GPT-5.2 Reflexion & Claude Sonnet 4.5 & 32.2 [25.8,39.1] & 89.5 [86.3,92.7] \\
 & GPT-5.4 & 34.5 [27.6,42.0] & 99.4 [98.9,99.8] \\
 & DeepSeek V4 Flash & 32.2 [25.9,39.1] & 80.4 [78.3,82.7] \\
\addlinespace
GPT-5-mini single-shot & Claude Sonnet 4.5 & 21.3 [14.9,28.2] & 74.8 [70.6,78.9] \\
 & GPT-5.4 & 25.9 [19.5,32.8] & 95.5 [93.7,97.0] \\
 & DeepSeek V4 Flash & 23.0 [16.7,29.3] & 36.6 [33.0,40.5] \\
\addlinespace
GPT-5-mini Reflexion & Claude Sonnet 4.5 & 28.2 [21.8,35.1] & 78.5 [74.7,82.4] \\
 & GPT-5.4 & 37.4 [29.9,44.8] & 97.5 [96.3,98.4] \\
 & DeepSeek V4 Flash & 35.6 [28.2,43.1] & 41.7 [38.2,45.9] \\
\addlinespace
Gemma 4 31B single-shot & Claude Sonnet 4.5 & 12.6 [8.0,17.8] & 67.9 [61.9,73.4] \\
 & GPT-5.4 & 14.9 [9.8,20.7] & 90.5 [87.2,93.6] \\
 & DeepSeek V4 Flash & 14.4 [9.8,19.5] & 35.2 [29.9,40.1] \\
\addlinespace
Gemma 4 31B Reflexion & Claude Sonnet 4.5 & 19.5 [13.8,25.3] & 75.2 [70.8,79.4] \\
 & GPT-5.4 & 23.6 [17.8,29.9] & 92.0 [89.9,93.9] \\
 & DeepSeek V4 Flash & 21.8 [16.1,28.2] & 38.7 [34.3,43.6] \\
\addlinespace
GPT-OSS 20B single-shot & Claude Sonnet 4.5 & 5.8 [2.9,9.2] & 52.2 [46.0,59.0] \\
 & GPT-5.4 & 6.3 [3.4,9.8] & 81.7 [76.5,86.7] \\
 & DeepSeek V4 Flash & 9.2 [5.2,13.2] & 25.9 [20.9,31.4] \\
\addlinespace
GPT-OSS 20B Reflexion & Claude Sonnet 4.5 & 12.1 [7.5,17.2] & 63.0 [57.0,69.5] \\
 & GPT-5.4 & 12.6 [8.0,18.4] & 86.8 [83.8,90.1] \\
 & DeepSeek V4 Flash & 15.5 [10.3,21.3] & 33.7 [29.2,39.0] \\
\addlinespace
Gemini Code Assist & Claude Sonnet 4.5 & 7.5 [3.4,11.5] & 77.7 [69.4,86.5] \\
 & GPT-5.4 & 8.1 [4.0,12.6] & 98.1 [96.1,99.7] \\
 & DeepSeek V4 Flash & 8.6 [4.6,13.2] & 58.1 [50.1,66.8] \\
\addlinespace
\bottomrule
\end{tabular}}
\caption{Performance by evaluation judge under the balanced prompt. Each judge independently classifies the same fixed review comments. Point estimates use all 174 instances. The 95\% percentile confidence intervals use 1,000 instance-level bootstrap resamples (seed 42), with comments grouped by instance. KDDR measures detection of the labeled defect, and Usefulness Rate measures the percentage of beneficial comments. See Section~\ref{sec: cr_verifier} for metric definitions and the statistical protocol.}
\label{tab:judge_exact}
\end{table*}

Across judges, Reflexion has higher KDDR than single-shot for every backbone. The magnitude of Usefulness Rate varies more because judges differ in how often they classify comments as Noise.

% Full-width float: these metric tables are far wider than one ACL column.
\begin{table*}[t]
\centering
\scriptsize
\resizebox{\textwidth}{!}{%
\begin{tabular}{@{}llcc@{}}
\toprule
Backbone & Evaluation judge & $\Delta$KDDR [CI], $p$ & $\Delta$Usefulness Rate [CI], $p$ \\
\midrule
GPT-5.2 & Claude Sonnet 4.5 & +5.7 [-0.0,11.5], 0.083 & +7.8 [5.2,10.7], 0.001 \\
 & GPT-5.4 & +3.4 [-2.3,9.2], 0.336 & +0.3 [-0.2,0.9], 0.284 \\
 & DeepSeek V4 Flash & +2.9 [-3.4,9.2], 0.504 & +2.6 [-0.2,5.3], 0.066 \\
\addlinespace
GPT-5-mini & Claude Sonnet 4.5 & +6.9 [1.1,13.2], 0.036 & +3.7 [0.2,7.4], 0.101 \\
 & GPT-5.4 & +11.5 [4.6,18.4], 0.002 & +2.1 [0.4,3.9], 0.012 \\
 & DeepSeek V4 Flash & +12.6 [5.7,19.5], 0.004 & +5.1 [1.2,9.0], 0.032 \\
\addlinespace
Gemma 4 31B & Claude Sonnet 4.5 & +6.9 [1.7,12.1], 0.027 & +7.4 [2.5,12.1], 0.007 \\
 & GPT-5.4 & +8.6 [2.3,14.9], 0.006 & +1.5 [-1.8,4.9], 0.318 \\
 & DeepSeek V4 Flash & +7.5 [1.7,13.8], 0.016 & +3.5 [-1.8,9.2], 0.251 \\
\addlinespace
GPT-OSS 20B & Claude Sonnet 4.5 & +6.3 [1.7,11.5], 0.023 & +10.8 [3.9,17.2], 0.001 \\
 & GPT-5.4 & +6.3 [1.7,11.5], 0.030 & +5.2 [0.3,10.2], 0.032 \\
 & DeepSeek V4 Flash & +6.3 [0.6,12.6], 0.060 & +7.8 [2.3,13.5], 0.004 \\
\addlinespace
\bottomrule
\end{tabular}}
\caption{Paired Reflexion-minus-single-shot differences for the same backbone and evaluation judge. Positive differences favor Reflexion. Strategies are matched by \texttt{instance\_id}. Confidence intervals use 1,000 paired bootstrap resamples, and $p$-values use two-sided paired permutation tests with 1,000 permutations. Gemini Code Assist is excluded because it has no matched single-shot and Reflexion pair. KDDR and Usefulness Rate differences are in percentage points. See Section~\ref{sec: cr_verifier} for metric definitions and the statistical protocol.}
\label{tab:paired_all_metrics}
\end{table*}

\begin{table*}[t]
\centering
\small
\begin{tabular}{@{}llcccc@{}}
\toprule
 & & \multicolumn{2}{c}{Single-shot} & \multicolumn{2}{c}{Reflexion} \\
\cmidrule(lr){3-4}\cmidrule(l){5-6}
Evaluation judge & Prompt & KDDR & U & KDDR & U \\
\midrule
Claude Sonnet 4.5 & Balanced & 26.4 & 81.7 & 32.2 & 89.5 \\
 & Strict & 25.9 & 38.2 & 32.8 & 50.7 \\
 & Lenient & 27.0 & 44.0 & 32.2 & 52.7 \\
\addlinespace
GPT-5.4 & Balanced & 31.0 & 99.1 & 34.5 & 99.4 \\
 & Strict & 24.7 & 94.3 & 32.2 & 98.8 \\
 & Lenient & 23.6 & 35.4 & 31.6 & 53.0 \\
\addlinespace
DeepSeek V4 Flash & Balanced & 29.3 & 77.8 & 32.2 & 80.4 \\
 & Strict & 21.8 & 41.1 & 30.5 & 51.0 \\
 & Lenient & 26.4 & 15.4 & 29.3 & 43.8 \\
\bottomrule
\end{tabular}
\caption{Prompt robustness for fixed GPT-5.2 single-shot and Reflexion outputs on all 174 verified instances. Each judge classifies the same comments under balanced, strict, and lenient evaluator prompts. KDDR and Usefulness Rate (U) are percentages. Higher is better.}
\label{tab:prompt_robustness}
\end{table*}

\subsection{KDDR by Taxonomy Group}

\begin{figure}[!htb]
     \centering
     \includegraphics[width=\linewidth]{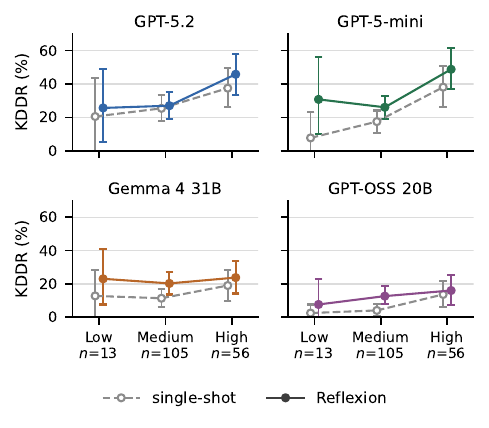}
     \caption{KDDR by defect severity for each backbone. Points average the three judges under the balanced prompt, and bars are 95\% bootstrap confidence intervals. The number of instances per severity level is printed on the axis. Table~\ref{tab:kddr_taxonomy} reports the same numbers for every category and impact group.}
     \label{fig:kddr_severity}
\end{figure}

Table~\ref{tab:kddr_taxonomy} reports KDDR for every taxonomy group, and Figure~\ref{fig:kddr_severity} plots the severity groups with confidence intervals.

% Granular taxonomy breakdown backing Section 5.3. Generated by
% paper_viz/taxonomy/generate_taxonomy_artifacts.py; n is printed so that the
% four groups with n<10 cannot be mistaken for evidence.
\begin{table*}[t]
\centering
\small
\begin{tabular}{llrcccc}
\toprule
\textbf{Dimension} & \textbf{Group} & \textbf{$n$} & \textbf{GPT-5.2} & \textbf{GPT-5-mini} & \textbf{Gemma 4 31B} & \textbf{GPT-OSS 20B} \\
\midrule
Category & Structural & 144 & 31.5 $\to$ 35.9 & 25.9 $\to$ 34.3 & 14.1 $\to$ 22.9 & 7.4 $\to$ 14.1 \\
 & IIS & 19 & 21.1 $\to$ 29.8 & 17.5 $\to$ 36.8 & 21.1 $\to$ 24.6 & 8.8 $\to$ 14.0 \\
 & RFF$^{\dagger}$ & 7 & 0.0 $\to$ 0.0 & 0.0 $\to$ 14.3 & 0.0 $\to$ 0.0 & 0.0 $\to$ 4.8 \\
 & Data$^{\dagger}$ & 3 & 33.3 $\to$ 0.0 & 0.0 $\to$ 33.3 & 0.0 $\to$ 0.0 & 0.0 $\to$ 0.0 \\
 & Memory$^{\dagger}$ & 1 & 0.0 $\to$ 0.0 & 0.0 $\to$ 33.3 & 0.0 $\to$ 0.0 & 0.0 $\to$ 0.0 \\
\midrule
Impact & Functional Suitability & 105 & 28.9 $\to$ 29.5 & 20.3 $\to$ 30.5 & 13.7 $\to$ 19.7 & 6.7 $\to$ 13.0 \\
 & Reliability & 41 & 35.8 $\to$ 39.8 & 31.7 $\to$ 38.2 & 14.6 $\to$ 22.0 & 10.6 $\to$ 17.1 \\
 & Compatibility & 13 & 12.8 $\to$ 43.6 & 33.3 $\to$ 43.6 & 17.9 $\to$ 33.3 & 5.1 $\to$ 12.8 \\
 & Usability & 13 & 20.5 $\to$ 25.6 & 7.7 $\to$ 30.8 & 12.8 $\to$ 23.1 & 2.6 $\to$ 7.7 \\
 & Perf. Efficiency$^{\dagger}$ & 2 & 50.0 $\to$ 50.0 & 50.0 $\to$ 66.7 & 0.0 $\to$ 33.3 & 0.0 $\to$ 0.0 \\
\midrule
Severity & Low & 13 & 20.5 $\to$ 25.6 & 7.7 $\to$ 30.8 & 12.8 $\to$ 23.1 & 2.6 $\to$ 7.7 \\
 & Medium & 105 & 25.4 $\to$ 27.0 & 17.5 $\to$ 26.0 & 11.4 $\to$ 20.3 & 4.1 $\to$ 12.7 \\
 & High & 56 & 37.5 $\to$ 45.8 & 38.1 $\to$ 48.8 & 19.0 $\to$ 23.8 & 13.7 $\to$ 16.1 \\
\bottomrule
\end{tabular}
\caption{KDDR (\%) for every taxonomy group, written as single-shot $\to$ Reflexion. Each value averages the three judges under the balanced prompt. $n$ is the number of \texttt{CR-Bench-Verified} instances in the group. $^{\dagger}$Groups with fewer than 10 instances are listed for completeness only. Their confidence intervals are tens of percentage points wide, so we draw no conclusions from them. The dimensions overlap: 12 of the 13 Usability instances are also the Low-severity instances, and Reliability is mostly high severity (26 High, 15 Medium, 0 Low). Differences across impact groups therefore partly reflect severity. Figure~\ref{fig:kddr_severity} plots the severity rows with confidence intervals. Figure~\ref{fig:kddr_taxonomy} plots every row. RFF = Requirements, Features and Functionality, IIS = Interface, Integration and System.}
\label{tab:kddr_taxonomy}
\end{table*}

\end{document}